\documentclass[11pt]{article}
\usepackage{amsfonts}
\usepackage{amsmath}
\usepackage{amssymb}
\usepackage{mathrsfs}
\usepackage[T1]{fontenc}
\usepackage{dsfont}
\usepackage{indentfirst}
\usepackage{color}
\usepackage{nicefrac}
\usepackage{esdiff}
\usepackage{nicefrac}
\usepackage{booktabs}
\usepackage{array}
\usepackage{paralist}
\usepackage{cite}
\numberwithin{equation}{section}
\usepackage{graphicx}
\usepackage{makeidx}
\usepackage{fancyhdr} 
\newcommand{\tr}{\mathrm{tr}\:}
\newcommand{\D}{\mathrm{d}}

\begin{document}
\title{\bf Higher charges and regularized quantum trace identities in su(1,1) Landau-Lifshitz model}

\author{ {\bf A. Melikyan}$\,^{1}$, {\bf A. Pinzul}$\,^{1}$, {\bf G. Weber}$\,^{2}$ \thanks{\tt amelik@gmail.com, apinzul@unb.br, weber@fma.if.usp.br}\\
\\
$^1$ \sl{Instituto de F\'{\i}sica}\\
\sl{Universidade de Bras\'{\i}lia, 70910-900, Bras\'{\i}lia, DF, Brasil}\\
\sl{and}\\
\sl{{International Center of Condensed Matter Physics} }\\
\sl{C.P. 04667, Brasilia, DF, Brazil} \\
\\
$^2$ \sl{Instituto de F\'{\i}sica}\\
\sl{Universidade de S\~{a}o Paulo, 05315-970, S\~{a}o Paulo, SP, Brasil}\\
}
\date{}
\maketitle

\begin{abstract}
	We solve the operator ordering problem for the quantum continuous integrable su(1,1) Landau-Lifshitz model, and give a prescription to obtain the quantum trace identities, and the spectrum for the higher-order local charges. We also show that this method, based on operator regularization and renormalization, which guarantees quantum integrability, as well as the construction of self-adjoint extensions, can be used as an alternative to the discretization procedure, and unlike the latter, is based only on integrable representations.
\end{abstract}

\newpage
\tableofcontents

\section{Introduction}

{It is well-known that the continuous integrable models are much harder to quantize than the discrete ones. On the other hand, the discretization of a given continuous model may not be known, and the standard techniques to obtain one lead to quite complicated results \cite {Izergin:1982ry,tarasov1983lhi,tarasov1984lhi, Korepin:1997bk, kundu1992ciq, Kundu:1982wf}. One immediately faces several problems working directly with a continuous integrable model. The main difficulty is the problem of singularities in the operator product in the quantum theory, and, as a result, the operator ordering problem. There are several known continuous models that work out without any serious problems, e.g., the non-linear Schrödinger and Thirring models \cite{Korepin:1997bk}. However, as it was emphasized in the previous work \cite{Melikyan:2008ab}, the derivation of the integrability in those models is not mathematically rigorous, although the formal manipulations with singular objects lead to the correct results. The mathematically strict method to deal with such models was first obtained in \cite{Melikyan:2008ab}, where on the example of the highly singular $su(1,1)$ Landau-Lifshitz (LL) model it was shown that not only one cannot do formal manipulations with singular potentials of the $\delta ^{\prime \prime }(x)$ type, but it is necessary to construct self-adjoint extensions and work with regularized operator product. The physical consequences of such operations are significant. For example, the thermodynamical properties of the model considered in \cite{Melikyan:2010bi} turn out to be quite non-trivial, and allowed us to show the stability of the system. Another more important consequence concerns the widely accepted representation of the $n$-particle wave-function in the ``factorizable'' form, e.g., for $n=2$:
\begin{equation}
	\left\vert p_{1}p_{2}\right\rangle =\int \int dx_{1}dx_{2}\left[ e^{p_{1}x_{1}+p_{2}x_{2}}+\left( S\right) e^{p_{1}x_{2}+p_{2}x_{1}}\right] \label{2pp}
\end{equation}
which, as we showed, is not correct in general. It is certainly not correct for the Landau-Lifshitz model, or the fermionic $AAF$ model \cite{Alday:2005jm}. It is also not correct, for the same reason, for the principal chiral model, quantization and integrability of which was considered in \cite{Faddeev:1985qu,Klose:2006dd,Das:2007tb}, even though, in this case one does not encounter singular potentials such as $\delta ^{\prime \prime }(x)$. Nevertheless, even in this usual $\delta (x)$ interaction it is necessary to construct self-adjoint extensions. Thus, the continuous integrable models should be treated with more mathematical care.\footnote{ These different integrable models arise in different sub-sectors of the $ AdS_{5}\times S^{5}$ string and correspondingly require different constructions of self-adjoint extensions. The interesting question of constructing a larger space to accommodate the full string sector will be investigated in a separate publication.} }

In \cite{Melikyan:2008ab} we have worked out this method in details on the example of the quantum Hamiltonian for the $su(1,1)$ Landau-Lifshitz model. The natural question is how to extend this method to higher order charges. In this article we address this issue, and outline the procedure to obtain the quantum trace identities, which should be applicable for a wide range of continuous integrable models. Let us remind, that, for example, even for the much simpler $NLS$ model, obtaining the quantum trace identities is not a simple task \cite{Korepin:1997bk, Izergin:1982ry,Izergin1981d,kundu1992ciq,tarasov1983lhi,tarasov1984lhi, Faddeev:1992xa, Boos2003b}. In particular, the widely accepted notion that the normal ordering with respect to the bosonic fields $ \psi(x) $ and $\psi ^{+}(x)$ solves the operator ordering problem, is not quite correct. In fact, although there are no problems for the first three charges following from the quantum trace identities, the forth charge does not correspond to the usual normal ordering \cite{Korepin:1997bk,Davies:1990fk,Izergin1981c}. For the Landau-Lifshitz model, the situation is more complex, because now there are three fields $\vec{S}$ subject to be constrained on a hyperboloid. It is not clear \emph{a priori} how to choose consistent operator ordering in this case. In the preliminary paper of Sklyanin \cite{Sklyanin:1988s1}, the attempt was made to resolve the constraint, and then use the normal ordering for the remaining two operators. However, this procedure faces significant difficulties on the later stage. For example, in this formalism, it was not possible to obtain the quantum Hamiltonian in terms of the unconstrained fields, while in terms of the $\vec{S}$ fields the Hamiltonian, as we had shown in \cite{Melikyan:2008ab} has a quite simple form. 

This issue becomes even more important for other, more complicated continuous models, where resolving the constraints may not be technically possible. Thus, the first problem we address in this paper is how to work with operator product in this case. This is done with a special operator regularization and renormalization procedure. Namely,  we show that it leads to the intertwining relation, which is fundamental for quantum integrability and, therefore, quantum trace identities. The regularization and renormalization of operators that we use essentially differ from the one used initially by Sklyanin. The latter was used to only satisfy the intertwining relation, thus, regularizing the singularities only of a special type. However, as we show below, it is not enough to correctly quantize the model, and, therefore, we introduce a more general operator regularization and renormalization, which takes care of all singularities, and which has a number of interesting properties and consequences. In particular, the regularized and renormalized operator algebra turns out to be highly non-local even for the \emph{isotropic} LL-model,\footnote{ Let us note, that such non-local algebras (the \emph{Sklyanin algebras}) originally arose from the \emph{anisotropic} LL model \cite{Sklyanin:1988s1,Sklyanin:1982tf,Sklyanin:1983ig}. Here we show that it is necessary to introduce an analogue of the Sklyanin algebra even for the isotropic case.} and as a result, the subsequent analysis becomes much more involved. Here we also note, that the standard derivation of the quantum trace identities for continuous models does in fact use \cite{Izergin1981c,Korepin:1997bk}, as an intermediate step, the discretization of the model, which, as we emphasized above, is not always an obvious procedure. We discuss the steps to avoid this difficulty and at no point introduce any discretization. Moreover, we show that unlike the discretization procedure, which should utilize non-integrable representations, in our method we use only integrable representations, and verify this result for the next higher-order charge.

The second step we consider is the necessary self-adjointness check of each operator, associated with the conserved quantity, which requires the construction of self-adjoint extensions. This leads to mathematically correct description in the quantum mechanical formalism \cite {Melikyan:2008ab}, and which clearly shows that the intuitively obvious expressions of the type (\ref{2pp}) are not always correct. We demonstrate this check for the Hamiltonian, and the next non-trivial cubic operator.

The third step in our method utilizes the $\mathcal{M}$-operator formalism, which establishes a connection between our method and the quantum inverse scattering method, and which makes it possible to actually derive the spectrum for each operator without, as it was initially done by Sklyanin, guessing it and using its classical limit. We again demonstrate this point for the Hamiltonian and the next non-trivial charge. In the process, we also discuss a number of important related points. In particular, the constraint regularization and renormalization are carefully considered, and the quantum analogue of it is derived. We show that after all regularizations are removed, the resulting Casimir operator corresponds to integrable representations, consistent with self-adjointness, and boundary conditions.

We also present the classical inverse scattering method analysis for the $ su(1,1)$ LL-model, on which we subsequently rely during the analysis of the quantum model. Finally, by comparing our results to the classical ones, and using the symmetry of the model, we argue that any higher-order quantum conserved charge can be written in terms of the three fundamental charges: the Casimir operator, the quantum Hamiltonian, and the next non-trivial conserved local charge, which essentially solves the problem of quantum trace identities for the $su(1,1)$ continuous LL-model.

The paper is organized as follows. In Section \textbf{\ref{overviewofLL}} we give the main relevant expressions and definitions related to the quantization of the $ su(1,1)$ $LL$-model. In particular, in Section \textbf{\ref{ismclassic}} we give all essential results of the classical inverse scattering method. In Sections \textbf{\ref{overviewofLLquant}} and \textbf{\ref{overviewofLLMoperator}} we overview the quantization procedure, and discuss the Sklyanin product, as well as the construction of self-adjoint extensions and the hierarchy of $\mathcal{M}_m$-operators. The main results are presented in Section \textbf{\ref{operatorregularization}} where we introduce the operator regularization, discuss the resulting algebraic structure and show the quantum integrability. We also derive the quantum fundamental charges, their spectrum, and establish a connection to the quantum inverse scattering method. Finally, in Section  \textbf{\ref{sec:self}} we check the self-adjointness of the higher-order fundamental charge. In conclusion, Section \textbf{\ref{conclusion}}, we discuss the obtained results, and propose further directions to develop the ideas presented in this paper for other, more interesting continuous integrable models.

\section{Overview of the Landau-Lifshitz model}

\label{overviewofLL}

In this section we make a short review, following \cite{Sklyanin:1988s1,fogedby1980sci, fogedby1980msp, fogedby1980sam}, of the Landau-Lifshitz model in the context of the inverse scattering method and discuss the arising difficulties upon the attempt to obtain algebraic expressions for the conserved charges in the quantum theory.

The most general anisotropic Landau-Lifshitz model is described in the classical case by the following Hamiltonian (see Section \textbf{\ref{ismclassic}} for more details):
\begin{equation}
	H = \frac{\varepsilon}{2} \int dx \: \left\{ - \partial_x {S^i}	\partial_x {S_i} + 4 \gamma^2 \left[\left(S^3\right)^2 - 1 \right] \right\} \label{anisotropichamiltonian}
\end{equation}
where $\varepsilon = \pm 1$ and the fields $S^{i}$ $(i =1,2,3)$ define a vector in three-dimensional space, $\vec{S} = (S^1,S^2,S^3)$ with the scalar product defined by:
\begin{equation}
	\label{scalarproduct} S^i S_i = -\varepsilon \left(S^1\right)^2 -\varepsilon \left(S^2\right)^2 + \left(S^3\right)^2 = 1
\end{equation}
It is convenient to introduce $S^{\pm }=S^{1}\pm iS^{2}$, in terms of which the Poisson structure has the form:
\begin{align}
	\left\{ S^{3}(x),S^{\pm }(y)\right\} & =\pm iS^{\pm }(x)\delta (x-y) \notag \\
	& \label{poissonstructure} \\
	\left\{ S^{-}(x),S^{+}(y)\right\} & =2i\varepsilon S^{3}(x)\delta (x-y) \notag
\end{align}

The sign of $\varepsilon$ determines whether the model is defined on the $ su(2)$ sphere $(\varepsilon = -1)$ or on the $su(1,1)$ hyperboloid $ (\varepsilon = 1)$. As noted in \cite{Sklyanin:1988s1}, it is only possible to construct physically meaningful states over the ferromagnetic vacuum for the hyperboloid case. The role of the parameter $\gamma$ is to introduce anisotropy in the model and, as explained in \cite{Sklyanin:1988s1}, the isotropic case $(\gamma = 0)$ should be treated separately from the more general anisotropic $(\gamma \neq 0)$ one. Unlike the former case, where the standard inverse scattering method works out throughout, that is, the Yang-Baxter and bilinear equations are satisfied with the usual choice of the $R$-matrix and yield the commuting family of operators, the latter requires a modification of the monodromy matrix and the spin operator algebra (giving rise to the quadratic {\em Sklyanin} algebra \cite{Sklyanin:1982tf,Sklyanin:1983ig}), so that the intertwining relations have a solution with the $R$-matrix of the $XYZ$-model. In this article we focus only on the isotropic $ su(1,1)$ case.

\subsection{Classical inverse scattering method} \label{ismclassic}

We start our brief review of the classical inverse scattering method with fixing the notations. This will allow us to treat both $su(2)$ and $su(1,1)$ models simultaneously (see below the crucial difference between these models in the quantum case). Introducing the metric
\begin{equation}
	\eta_{ij}=diag(-\varepsilon , -\varepsilon , 1) \label{metric}
\end{equation}
where again $\varepsilon = -1$ corresponds to the $su(2)$ case and $\varepsilon = 1$ corresponds to the $su(1,1)$ one, we can write the equations of motion for the spin field, $S^i(x)$, in the usual form:
\begin{equation}
	\frac{\D S^i}{\D t} =\epsilon^{ijk}S_j
	\partial^2 S_k \label{eom}
\end{equation}
where $\epsilon^{123}=1$, $
\partial := \frac{
\partial}{
\partial x}$ and we used the metric (\ref{metric}) to lower indices, i.e., $S_i = \eta_{ij}S^j$. It is clear that the dynamics is consistent with the constraint
\begin{equation}
	\eta_{ij}S^i S^j = 1 \label{constraint}
\end{equation}
The equation (\ref{eom}) plus the constraint (\ref{constraint}) define the classical dynamics of the $su(2)$ and $su(1,1)$ models.

It is easily verified that (\ref{eom}) can be obtained as the canonical equations using the following Hamiltonian and Poisson structure (c.f., Eqs.(\ref{anisotropichamiltonian}-\ref{poissonstructure})):
\begin{align}
	H = -\frac{\varepsilon}{2}\int\D x\ \eta_{ij}
	\partial S^i(x)
	\partial S^j(x) \label{Ham}\\
	\{ S^i(x), S^j(y)\} = \varepsilon \epsilon^{ijk} S_k(y) \delta(x-y) \label{Pois}
\end{align}
Now we present a brief account on how to solve (\ref{eom}) using the inverse scattering method. Here we mostly follow \cite{fogedby1980sci, fogedby1980msp, fogedby1980sam}, generalizing it to incorporate $su(1,1)$ case. With the help of $\sigma$-matrices
\begin{align}
	[\sigma^i , \sigma^j] &= -2 i \varepsilon \epsilon^{ijk}\sigma_k \\
	\{ \sigma^i , \sigma^j\} &= 2\eta^{ij}
\end{align}
one can introduce a matrix-valued field $\mathbf{S}$:
\begin{equation}
	\mathbf{S}:= S^i \sigma_i
\end{equation}
It can be easily checked that Eqs. (\ref{eom}) and (\ref{constraint}) can be re-written as follows:
\begin{align}
	\frac{\D \mathbf{S}}{\D t} &= \varepsilon \frac{i}{2}[\mathbf{S} ,
	\partial^2\mathbf{S}] \\
	\mathbf{S}^2 &= 1
\end{align}
To pass to the auxiliary spectral problem, we need to construct a Lax pair. By a direct calculation, it is verified that the following choice
\begin{align}
	\mathbf{L} = \mathbf{S}
	\partial \mbox{\ \ and\ \ } \mathbf{M} = -\varepsilon (2\mathbf{S}
	\partial^2 +
	\partial \mathbf{S}
	\partial) \label{LMpara}
\end{align}
makes (\ref{eom}) and (\ref{constraint}) equivalent to:

i) the auxiliary spectral problem:
\begin{equation}
i\mathbf{L}\psi = \lambda \psi \label{speceq}
\end{equation}

ii) the time evolution of $\psi$, given by:
\begin{equation}
i\frac{\D \psi}{\D t} = \mathbf{M}\psi \label{evolution}
\end{equation}
As a result, the crucial property, that $\lambda$ is a constant of motion, can be proven.

With the help of the constraint (\ref{constraint}) the spectral equation (\ref{speceq}) can be brought to a more convenient form\footnote{Thus, the classical $L^{\textsl{(cl)}}$-operator, which enters the equation $\partial \psi= L^{\textsl{(cl)}}\psi$, has the form:
\begin{equation}
L^{\textsl{(cl)}} = -i\lambda \left(
	\begin{array}{cc}
		S^3(x) & \,\, iS^-(x) \\
		iS^+(x) & -S^3(x)
	\end{array}
\right) \notag
\end{equation}
}
\begin{equation}
	i
	\partial \psi = \lambda\mathbf{S}\psi \label{inverse}
\end{equation}
It is important to notice that this equation as well as the asymptotic for $\mathbf{S}$: $\mathbf{S}(x)\underset{|x|\rightarrow\infty}{\longrightarrow} \sigma^3$ do not depend on $\varepsilon$, i.e., both sets of invariants, for $su(2)$ and $su(1,1)$, will be exactly the same if written in a covariant form, i.e., using the metric (\ref{metric}) to contract indices (see below).

Equations (\ref{evolution}) and (\ref{inverse}) are the starting point of the classical inverse scattering method. Here we only sketch the main steps leading to the construction of the infinite set of invariants.

One starts with (\ref{inverse}) to construct two Jost solutions, $F(x,\lambda)$ and $G(x,\lambda)$, which satisfy the following asymptotical conditions:
\begin{align}
	F(x,\lambda) \underset{x\rightarrow +\infty}{\longrightarrow} e^{-i\lambda\sigma^3 x}\\
	G(x,\lambda) \underset{x\rightarrow -\infty}{\longrightarrow} e^{-i\lambda\sigma^3 x}
\end{align}
These solutions depend on $S^i(x)$ through $\mathbf{S}(x)$ which plays the role of the potential in (\ref{inverse}). Then the solution for the spin-field, $S^i(x)$, is obtained by solving Gelfand-Marchenko equation, which though being integral equation, still is simpler than the original Landau-Lifshitz non-linear problem.

The time dependence of the solutions is controlled by (\ref{evolution}). It is important to notice that $\varepsilon$ in the definition of $\mathbf{M}$-operator can be eliminated by the time redefinition: $t\rightarrow \varepsilon t$. Then the evolution of both, $su(2)$ and $su(1,1)$, models is the same in terms of the new time. In particular, the invariants will be the same. Introducing the transition matrix, $T({\lambda})$, which connects two Jost solutions $$ G(x,\lambda) = F(x,\lambda) T({\lambda}) $$ and using (\ref{evolution}) one can easily find the time dependence of $T({\lambda})$:
\begin{align}
	T(\lambda, t) = e^{-2i\varepsilon \lambda^2 \sigma^3 t} T(\lambda, 0) e^{2i\varepsilon \lambda^2 \sigma^3 t}
\end{align}
It is well known that the canonical action-angle variables are constructed out of the matrix elements of $T(\lambda, t)$, and, what is more important for us, this leads to an infinite set of conserved charges. In particular, the diagonal element of $T(\lambda, t)$, $a(\lambda, t)$, is independent of time. Then one can obtain all, local and non-local, conserved charges as coefficients of asymptotics for $Im \ln a(\lambda)$:
\begin{align}
	Im \ln a(\lambda) &= - \sum^{\infty}_{k=0} \lambda^{-k} A_k \ \ ,\ |\lambda|\longrightarrow \infty \label{loc}\\
	Im \ln a(\lambda) &= - \sum^{\infty}_{k=1} \lambda^{-k} B_k \ \ ,\ |\lambda|\longrightarrow 0 \label{nonloc}
\end{align}
where $A_k$ and $B_k$ are {\it global} charges given by some known expressions in terms of the action variables (see for the details \cite{fogedby1980sci, fogedby1980msp, fogedby1980sam}). It is important that these charges can be represented as integrals of the {\it local},\footnote{This should not be confused with local charges defined as conserved quantities that locally depend on a field $S^i (x)$ and a finite number of its spatial derivatives. In fact, it is well known that series (\ref{loc}) defines exactly such charges, while charges $b_k$, defined by (\ref{nonloc}), have the form of spatial integrals (except for $b_1$ that is just $S^3(x)-1$).} i.e., depending on $x$, quantities:
\begin{align}
	A_k = \int \D x\ a_k (x) \mbox{\ and\ }\ B_k = \int \D x\ b_k (x) \label{AB}
\end{align}
There is a well-established recursive procedure to obtain $a_k (x)$ and $b_k (x)$, e.g.:
\begin{align}
	a_0 (x) &= \frac{1}{4} \frac{S^2 (x)
	\partial S^1 (x) - S^1 (x)
	\partial S^2 (x)}{1+S^3 (x)} \\
	a_1 (x) &= \frac{1}{8} \eta_{ij}
	\partial S^i (x)
	\partial S^j (x)
\end{align}
While $a_0$ is proportional to the momentum, $a_1$ is easily recognized as the Hamiltonian density (\ref{Ham}) of the model. The higher classical charges could be obtained in the same vein.

Instead, we will proceed following \cite{Jevicki:1978yv}. As explained in \cite{Jevicki:1978yv}, in the classical theory an arbitrary $O(1,2)$ invariant\footnote{Actually, in \cite{Jevicki:1978yv} only the $O(3)$ case was considered. But as we already stressed, both, $su(2)$ and $su(1,1)$, models have exactly the same set of invariants if written in invariant form, which is the case for $\mathcal{C}$, $\mathcal{H}$ and $\mathcal{Q}_3$.} can be written in terms of three basis invariants and their space derivatives:
\begin{align}
	\mathcal{C}={S^i}{S_i};\quad \mathcal{H}(x)=
	\partial_x {S^i}
	\partial_x {S_i}; \quad \mathcal{Q}_{3}(x)= \epsilon^{ijk}S_{i}
	\partial_{x} S_{j}
	\partial_{x}^{2} S_{k} \label{invariants}
\end{align}
where $\mathcal{C}$, $\mathcal{H}(x)$, and $\mathcal{Q}_3(x)$ correspond to the densities of the Casimir, Hamiltonian and the next cubic conserved charges\footnote{For example, the quartic conserved charge $Q_4 = \int dx \left[
\partial_{x}^{2} {S^i}
\partial_{x}^2 {S_i} + \nicefrac{5}{4} \left(
\partial_x {S^i}
\partial_x {S_i} \right)^2 \right]$ can be written as a non-linear function of $H$ and $Q_3$ in the form: $Q_4= \int dx \left( 9H^2+\frac{(
\partial_{x}H)^{2}+Q_{3}^2}{2H}\right)$}. In Section \textbf{\ref{operatorregularization}} the appropriate quantum operators will be derived and, in particular, the Casimir operator will be shown to correspond, after removing regularization, to an integrable representation of the $su(1,1)$ algebra. We will also show that in the quantum theory, these fundamental invariants can be written as traces of regularized operators in a simple form.

\subsection{Quantization} \label{overviewofLLquant}

The quantization of the model involves promoting the classical $S^{i}$ fields to quantum operators which satisfy the canonical commutation relations obtained from (\ref{poissonstructure}) or (\ref{Pois}),
\begin{align}
	\left[ S^{3}(x);S^{\pm }(y)\right] & =\pm S^{\pm }(x)\delta (x-y) \notag \\
	& \label{canonicalcommutationrelations} \\
	\left[ S^{-}(x);S^{+}(y)\right] & =2\varepsilon S^{3}(x)\delta (x-y) \notag
\end{align}
and the conjugation conditions: $\left( S^{3}(x)\right) ^{\ast }=S^{3}(x)$ and $\left( S^{\pm }(x)\right) ^{\ast }=S^{\mp }(x)$.

The next step is to construct the representation of the algebra (\ref{canonicalcommutationrelations}) in the ferromagnetic vacuum:
\begin{equation}
	S^3(x) |0\rangle = \varepsilon |0\rangle \:, \quad S^-(x) |0\rangle = 0 \label{ferromagneticvacuum}
\end{equation}
One possible way is to proceed as in other continuous integrable models \cite {Korepin:1997bk} and introduce the vectors:
\begin{equation}
	|f_n\rangle = \int \prod_{i=1}^n dx_i \: f_n(x_1,\ldots, x_n ) \prod_{j=1}^n S^+(x_j) |0\rangle \label{representationofthealgebra}
\end{equation}
where the $f_n(x_1,\ldots,x_n)$ are continuous and sufficiently fast decreasing, symmetric functions of $\{x_n\}$ for the integral (\ref{representationofthealgebra}) to be well defined. So, the representation space is given by the linear span of the vectors $|f_n\rangle$, $ n=1,2,\ldots $. The scalar product in the $n$-particle subspace, $\langle g_n| f_n \rangle $, is positive only in the $su(1,1)$ case, it is indefinite in the $su(2)$ case \cite{Sklyanin:1988s1}. Hence, only in the non-compact case it is possible to construct physically meaningful representations of the algebra (\ref{canonicalcommutationrelations}) in the ferromagnetic vacuum. Thus, in order to obtain consistent result it is mandatory to set $ \varepsilon = 1$ in (\ref{anisotropichamiltonian}-\ref{ferromagneticvacuum}). We would like to stress that although there is no positively defined scalar product for the $su(2)$ case with respect to the ferromagnetic vacuum, it is, however, possible to construct a representation with $ \varepsilon = -1$ in a non-ferromagnetic vacuum \cite{albeverio1983frp}.

The quantum $\mathcal{L}$-operator for the isotropic case has the form:\footnote{The $\mathcal{L}(\lambda,x)$-operator (\ref{l-operator})  is the quantum version of the classical $\mathcal{L}^{\textsl{(cl)}}(\lambda,x)$-operator, which is obtained from $L^{\textsl{(cl)}}$ (see the footnote on page $\mathbf{6}$) by the following gauge transformation (we also take $\lambda \to \nicefrac{1}{\lambda}$): $\partial \phi =\mathcal{L}^{\textsl{(cl)}}(\lambda,x) \phi$; $\phi \equiv \Lambda \psi$; $\mathcal{L}^{\textsl{(cl)}}(\lambda,x) \equiv \Lambda L^{\textsl{(cl)}} \Lambda^{-1} $, where
$
\label{clasL1} \Lambda = \left(
	\begin{array}{cc}
		0 & 1 \\
		i & 0
	\end{array}
\right). \notag
$
This form of $\mathcal{L}^{\textsl{(cl)}}(\lambda,x)$ is more convenient in the quantum case because the quantum $R$-matrix (\ref{Rmatrix}) has the standard (as in $su(2)$ case) form. The fact that $\Lambda$ is not unitary is not a problem, as $\psi$ (or $\phi$) does not have any probabilistic interpretation as in the case of Quantum Mechanics.}
\begin{equation}
	\label{l-operator} \mathcal{L}(\lambda,x) = \frac{i}{\lambda}\left(
	\begin{array}{cc}
		S^3(x) & -S^+(x) \\
		S^-(x) & -S^3(x)
	\end{array}
	\right)
\end{equation}
Here $\lambda$ again is the spectral parameter. The monodromy matrix on the interval $[x_-,x_+]$ is defined as:
\begin{equation}
	T^{x_+}_{x_-}(\lambda) =P\,{\exp}\int_{x_-}^{x_+} dx \: \mathcal{ \ L}(\lambda,x) \equiv \left(
	\begin{array}{cc}
		A^{x_+}_{x_-}(\lambda) & B^{x_+}_{x_-}(\lambda) \\
		C^{x_+}_{x_-}(\lambda) & D^{x_+}_{x_-}(\lambda)
	\end{array}
	\right) \label{monmat}
\end{equation}
In this case, the bilinear relation,
\begin{equation}
	\label{bilinearrelation} R(\lambda_1 -\lambda_2) \overset{(1)}{T^{x_+}_{x_-}}(\lambda_1) \overset{(2)} {T^{x_+}_{x_-}}(\lambda_2) = \overset{(2)}{T^{x_+}_{x_-}}(\lambda_2)\overset{ (1)}{T^{x_+}_{x_-}}(\lambda_1) R(\lambda_1-\lambda_2)
\end{equation}
where we denote $\overset{(1)}{A} \equiv A \otimes \mathds{1}$ and $\overset{ (2)}{A} \equiv \mathds{1} \otimes A$, holds for the quantum $R$-matrix given by:\footnote{Everywhere below we use the standard $su(2)$ matrices $\sigma_{a}=({\mathds{1},\sigma_{i}})$}
\begin{equation}
	\label{Rmatrix} R(\lambda) = \sum^{3}_{a=0} w_a(\lambda) \sigma_a \otimes \sigma_a
\end{equation}
with $w_0(\lambda) = \lambda - \nicefrac{i}{2}$ and $w_{1,2,3} = - \nicefrac{i}{2}$, provided that
\begin{align}
	R(\lambda_1-\lambda_2) \left[ \overset{(1)}{\mathcal{L}}(\lambda_1,x) + \overset{(2)}{\mathcal{L}}(\lambda_2,x) + \overset{(1)}{\mathcal{L}} (\lambda_1,x) \circ \overset{(2)}{\mathcal{L}}(\lambda_2,x) \right] = \notag \\
	= \left[ \overset{(1)}{\mathcal{L}}(\lambda_1,x) + \overset{(2)}{\mathcal{L}} (\lambda_2,x) + \overset{(2)}{\mathcal{L}}(\lambda_2,x) \circ \overset{(1)}{ \mathcal{L}}(\lambda_1,x) \right] R(\lambda_1-\lambda_2) \label{bilinearrelationforL}
\end{align}
is satisfied by the quantum $\mathcal{L}$-operator. The $\circ$-product was originally introduced in \cite{Sklyanin:1988s1} in order to make sensible the ill-defined product of operators at the same point, and is defined for any pair of local fields $A(x)$ and $B(x)$ as:
\begin{equation}
	\label{sklyaninproduct} A(x) \circ B(x) \equiv \lim_{\Delta \to 0} \frac{1}{\Delta} \int _x^{x+\Delta} d\xi_1 \: \int _x^{x+\Delta} d\xi_2 \: A(\xi_1) B(\xi_2)
\end{equation}
Even though (\ref{sklyaninproduct}) is enough for (\ref{bilinearrelationforL}) to hold, it is not in general a good prescription for the regularized product of operators at the same point, as it takes care of only one type of singularity appearing in the (\ref{bilinearrelationforL}). A quick analysis of (\ref{sklyaninproduct}) shows that its action formally corresponds to dividing by $\delta(0)$ the product of operators at the same point. Namely, it is easy to see that the product of two operators satisfying the algebra (\ref{canonicalcommutationrelations}) is regularized by the product (\ref{sklyaninproduct}) in the following sense:
\begin{equation}
	\label{prototypeofrelationssklyaniproduct} \left[ S^3(x), S^+(x)\right] = S^+(x) \delta(0) \overset{(\ref{sklyaninproduct})}{\longrightarrow} \left[ S^3(x)\overset{\circ}{,} S^+(x) \right] = S^+(x)
\end{equation}
Before moving on, we note that only the second expression in (\ref{prototypeofrelationssklyaniproduct}) is the necessary algebraic structure to render (\ref{bilinearrelationforL}) true. But now it is clear that (\ref{sklyaninproduct}) is neither suitable for dealing with the product of regular operators, nor of operators plagued with more severe singularities. For instance, if one takes $A(x)=B(x)=\mathds{1}$, the product defined by (\ref{sklyaninproduct}) implies that $\mathds{1} \circ \mathds{1} = 0$. Hence, it is not really a regularization, but a very specific procedure which demands a special care; later in section \textbf{\ref{operatorregularization}}, we will show that this problem can be properly addressed by introducing regularized $S$-operators. We emphasize that the the validity of (\ref{bilinearrelation}) and, therefore the standard methods of the quantum inverse scattering method \cite {Sklyanin:1980ij,Korepin:1997bk,Faddeev:1982rn} relies on the product (\ref{sklyaninproduct}).

After passing to the infinite line limit, the operators in the monodromy matrix (\ref{monmat}) satisfy the standard commutation relations and the quantities $ {\tr} \left[T(\lambda)\right]$ form a family of commuting operators, which can be regarded as the integrals of motion of the LL-model, and the eigenfunctions of which take the form:
\begin{equation}
	|\lambda_1\ldots \lambda_n \rangle = \prod_{i=1}^n B(\lambda_i) |0\rangle
\end{equation}
with the corresponding eigenvalues given by:
\begin{equation}
	\sum_{i=1}^n \log \left[ A(\lambda_i) \right]
\end{equation}

Here, however, one faces an important problem absent in the classical counterpart and the lattice version, namely, the difficulty to extract from $ {\tr} \left[T(\lambda)\right]$ the commuting quantities expressed as integrals of local densities, as in (\ref{loc}-\ref{AB}) in the classical case. The reason behind this difficulty lies in the fact that the local charges contain operator products at the same point, thus making the integrals of motion ill-defined quantities. In particular, one cannot proceed as in the classical case and simply decompose the series:
\begin{equation}
	{\tr} \left[T(\lambda)\right] = \sum_{n=0}^{\infty} I_n \lambda^n
\end{equation}
to obtain the local charges $I_n$. Hence, the construction of the local integrals of motion poses a highly non-trivial problem in the continuous quantum theory, which, from the field theory point of view, corresponds to the renormalization procedure (see for example \cite{jackiw:1995}).

In \cite{Sklyanin:1988s1} only the first two charges: the number operator $ \mathcal{N}$ and the momentum operator $\mathcal{P}$ were properly constructed. To formulate them, it was necessary to introduce a new set of bosonic fields $\Psi_n(x)$, corresponding to $n$-particle clusters, so that they annihilate the ferromagnetic vacuum, $\Psi_n(x) |0\rangle = 0$, and satisfy the following algebra:
\begin{equation}
	\label{psialgebra} \left[ \Psi_m(x); \Psi_n^{\dagger}(y) \right] = \delta_{mn} \delta(x-y)
\end{equation}

In terms of such $\Psi$-fields, one can represent the $S$-operators as:
\begin{align}
	S^3(x) &= s_0^3 + \sum_{n=1}^{\infty} s^3_n \Psi_n^{\dagger}(x)\Psi_n(x) \notag \\
	S^+(x) &= s_0^+ \Psi_1^{\dagger}(x) + \sum_{n=1}^{\infty} s^+_n \Psi^{\dagger}_{n+1}(x) \Psi_n(x) \label{sintermsofpsi} \\
	S^-(x) &= s_0^+ \Psi_1(x) + \sum_{n=1}^{\infty} s^+_n \Psi^{\dagger}_{n}(x) \Psi_{n+1}(x) \notag
\end{align}
where $s_0^3 = 1$, $s^+_0=\sqrt{2}$ and $s^3_n = n$, $s^+_n = \sqrt{n(n+1)}$ for $n \geq 1$.

In terms of the $\Psi$-fields the number and momentum operator become:
\begin{align}
	\mathcal{N} &= \sum_{n=1}^{\infty} n \int dx \: \Psi^{\dagger}_n(x) \Psi_n(x) \label{psinumber} \\
	\mathcal{P} &= \sum_{n=1}^{\infty} \frac{i}{2} \int dx\: \left(
	\partial_x \Psi^{\dagger}_n(x) \Psi_n(x) - \Psi^{\dagger}_n(x)
	\partial_x \Psi_n(x) \right) \label{psimomentum}
\end{align}

However, the quantum Hamiltonian was not found, only the action of the local quantum-mechanical Hamiltonian on one- and two-particle states was guessed. More recently, in \cite{Melikyan:2008ab} it was shown that in order to obtain the local quantum Hamiltonian for the arbitrary $n$-particle sector, it is necessary to simultaneously regularize the ill-defined operator product and construct the self-adjoint extensions. The first step in this construction is to introduce a regularized quantum Hamiltonian, which in \cite{Melikyan:2008ab} was achieved by means of the split-point regularization method.
\begin{align}
	H_Q = \frac{1}{2} \lim_{\epsilon \to 0} \int dx\: dy\: F_{\epsilon}(x,y) &\left\{-
	\partial_x S^3(x)
	\partial_y S^3(y) +
	\partial_x S^+(x)
	\partial_y S^-(y) + \right. \notag \\
	&+ \left.
	\partial_x
	\partial_y \left[ S^3(x) \delta(x-y) \right] -
	\partial_x
	\partial_y \delta(x-y) \right\} \label{splitpointhamiltonian}
\end{align}
where $F_{\epsilon}(x,y)$ is an arbitrary smooth and sufficiently fast decreasing symmetric function of $(x,y)$, depending on some parameter $\epsilon$, so that the integration in (\ref{splitpointhamiltonian}) is well-defined and that in the limit of vanishing $\epsilon$ one has:
\begin{equation}
	\lim_{\epsilon \to 0} F_{\epsilon}(x,y) = \delta(x-y)
\end{equation}
It is clear from the definition of $H_Q$ that it annihilates the ferromagnetic vacuum:
\begin{equation}
	H_Q |0\rangle = 0
\end{equation}
It is also not difficult to compute the action of the Hamiltonian (\ref{splitpointhamiltonian}) on the general $n$-particle state (\ref{representationofthealgebra})
\begin{align}
	H_Q |f_n\rangle &= - \int d\mathbf{x} \: \left[ \triangle_2 f_n(\mathbf{x}) \right] \prod_{i=1}^n S^+(x_i) |0\rangle \label{splitpointhamiltonianonfn} \\
	&+ \sum_{i>j} \int \prod_{k\neq j} dx_k \: \left\{ \left[
	\partial_i f_n( \mathbf{x}) -
	\partial_j f_n(\mathbf{x}) \right]_{x_i = x_j + \epsilon}^{x_i=x_j-\epsilon} +
	\partial_i
	\partial_j f_n(\mathbf{x} )\vert_{x_i=x_j} \right\} \prod_{i=1}^n S^+(x_i)\vert_{x_i=x_j} |0\rangle \notag
\end{align}
where $\triangle_2 = \sum_{i=1}^n
\partial_i^2$ stands for the $n$ -dimensional Laplacian. Thus for $|f_n\rangle$ to be an eigenstate of the Hamiltonian $H_Q$, one must require the following matching conditions:
\begin{equation}
	\label{matchingconditiossplitpointhamiltonian} \left[
	\partial_i f_n(\mathbf{x}) -
	\partial_j f_n(\mathbf{x}) \right]_{x_i = x_j + \epsilon}^{x_i=x_j-\epsilon} +
	\partial_i
	\partial_j f_n(\mathbf{x} )\vert_{x_i=x_j} = 0\:, \quad \forall i > j
\end{equation}
Thus, it follows from (\ref{splitpointhamiltonianonfn}) that
\begin{equation}
	\label{splitpointhamiltonianenergy} - \triangle_2 f_n(\mathbf{x}) = E_n f_n(\mathbf{x}) \Leftrightarrow H_Q|f_n\rangle = E_n |f_n\rangle
\end{equation}
where $E_n$ is the energy of the $n$-particle state. We would also like to point out before moving on to the construction of the self-adjoint extensions that, as explained in \cite{Melikyan:2008ab}, the matching conditions (\ref{matchingconditiossplitpointhamiltonian}) lead to the factorization property of the $S$-matrix.

For simplicity, here we will only consider the two-particle case and refer the interested reader to \cite{Melikyan:2008ab} for the details concerning the construction of the self-adjoint extensions for the general $ n $-particle sector. First, one introduces the vector space $V$ generated by vectors of the form
\begin{equation}
	\Psi = \left(
	\begin{array}{c}
		f_1(x) \\
		f_2(x,y)
	\end{array}
	\right) \label{vectors}
\end{equation}
where $f_1(x) \in \mathscr{L}^2(\mathbb{R},dx)$, $f_2(x,y) \in \mathscr{L} ^2( \mathbb{R}^2 \setminus \{x=y\}, dx\:dy)$ and $f_2(x,y)\vert_{x=y} = f_1(x) \in \mathscr{C}^0(\mathbb{R})$. A scalar product in $V$ is defined for any pair $\Psi,\Phi \in V$ as
\begin{equation}
	\label{scalarproductinV} \langle \Phi| \Psi \rangle = \frac{1}{2} \int dx\: g_1^*(x) f_1(x) + \int_{x\neq y} dx\: dy\: g_2^*(x,y) f(x,y)
\end{equation}

The operator $\hat{H}: V \to V$ is defined by the following two conditions:
\begin{enumerate}
	\item The action of $\hat{H}$ on $f_2(x,y)$ is simply the Laplacian $ -\triangle_2$ everywhere in $\mathbb{R}^2 \setminus \{x=y\}$, i.e.,
	\begin{equation}
		\hat{H} \Psi = \left(
		\begin{array}{c}
			\hat{h} f_1(x) \\
			-\triangle_2 f_2(x,y)
		\end{array}
		\right)
	\end{equation}
	
	\item $\hat{H}$ is Hermitian in $V$ with respect to the scalar product (\ref{scalarproductinV}), i.e.,
	\begin{equation}
		\langle \hat{H} \Phi | \Psi \rangle = \langle \Phi | \hat{H} \Psi \rangle
	\end{equation}
\end{enumerate}

These conditions together with the closure of $V$ under $\hat{H}$, i.e., $ \hat{H} \Psi \in V$, fix the form of $\hat{h}$ and we obtain the hamiltonian action:
\begin{equation}
	H_Q |f_2\rangle = \left(
	\begin{array}{c}
		2\left(
		\partial_x -
		\partial_y\right) f_2(x,y)\vert_{y=x-\epsilon}^{y=x+\epsilon} -
		\partial_x^2 f_2(x,x) \\
		- \triangle_2 f_2(x,y)
	\end{array}
	\right)
\end{equation}
which coincides with the formula previously guessed by Sklyanin in \cite{Sklyanin:1988s1} and agrees with (\ref{matchingconditiossplitpointhamiltonian}).

\subsection{$\mathcal{M}_{m}$-operator hierarchy} \label{overviewofLLMoperator}

As is well-known, in a classical integrable model the time evolution of a field $\Phi (x,t)$: $
\partial _{t}\Phi (x,t)=\left\{ I_{m},\Phi (x,t)\right\}$, where an arbitrary conserved charge $I_{m}$ is chosen as the Hamiltonian, is equivalent to the hierarchy of $\mathcal{M}_{m}^{\textsl{(cl)}}(\lambda,x)$ operators, satisfying the compatibility condition\footnote{We omit everywhere the explicit dependence on time.} \cite{Korepin:1997bk}:
\begin{equation}
	\label{zerocurvaturecondition}
	\partial_t \mathcal{L}^{\textsl{(cl)}}(\lambda,x) =
	\partial_x \mathcal{M}_{m}^{\textsl{(cl)}}(\lambda,x) + \left[ \mathcal{M}_{m}^{\textsl{(cl)}}(\lambda,x), \mathcal{L}^{\textsl{(cl)}}(\lambda,x) \right]
\end{equation}
where $\mathcal{L}^{\textsl{(cl)}}(\lambda,x)$ is the classical limit of the quantum $\mathcal{L}(\lambda,x)$-operator (\ref{l-operator}) (see the footnote on page $\mathbf{8}$). For example, the standard $(\mathbf{L},\mathbf{M})$-pair of the classical theory (\ref{LMpara}) can be obtained from the pair $(\mathcal{L}^{\textsl{(cl)}}(\lambda,x),\mathcal{M}_{2}^{\textsl{(cl)}}(\lambda,x))$.

For a system defined on the interval $ [x_{1},x_{2}]$ with periodic boundary conditions, the solution has the form \cite{Faddeev:1987ph,Sklyanin:1980ij}:
\begin{equation}
	\left\{ {\tr}T_{x_{1}}^{x_{2}}(\lambda ),\mathcal{L}^{\textsl{(cl)}}(\mu ,x)\right\} =
	\partial _{x}\mathcal{M}_{x_{1}}^{x_{2}}(x;\lambda ,\mu )+\left[ \mathcal{M} _{x_{1}}^{x_{2}}(x;\lambda ,\mu ),\mathcal{L}^{\textsl{(cl)}}(x,\mu )\right]
\end{equation}
where
\begin{equation}
	\mathcal{M}_{x_{1}}^{x_{2}}(x;\lambda ,\mu )=\overset{(1)}{\tr} \left( \overset{(1)}{T_{x}^{x_{2}}}(\lambda )r(\lambda -\mu )\overset{(1)}{ T_{x_{1}}^{x}}(\lambda )\right) \label{classicalMmatrixrelation}
\end{equation}
and $r(\lambda -\mu )$ is the classical $r$-matrix (see also (\ref{rmat2}) below) and the operation $\overset{(1)}{\tr}$ is the trace operator in the first space in $\mathbb{C} ^{2}\otimes \mathbb{C}^{2}$:
\begin{equation}
	\overset{(1)}{\tr}\left( A\otimes B\right) =\overset{(1)}{\tr}\left( \overset{(1)}{A}\cdot \overset{(2)}{B}\right) =({\tr} A)B \notag
\end{equation}
Now, in the infinite interval limit the $\mathcal{M} _m^{\textsl{(cl)}}(\lambda,x)$ operator corresponding to the conserved charges $I_m$ can be obtained from the series $\mathcal{M}_{-\infty}^{+\infty}(x;\lambda ,\mu )= \sum_{m} \mu^{-m}\mathcal{M}_m^{\textsl{(cl)}}(\lambda,x)$.

In this paper we will be more interested in the quantum operators $\mathcal{M}_m(\lambda,x)$, which describe the temporal evolution of the quantum transition matrix $ T^{x_2}_{x_1}(\lambda)$. For the simpler non-linear Schrödinger model the details and explicit relations can be found in \cite{Sklyanin:1980ij}. For the LL-model, on the other hand, due to the difficulties related to obtaining the local integrals of motion it was only possible obtain the quantum $\mathcal{M}_m(\lambda,x)$-operators:
\begin{equation}
	i \left[ I_n, \mathcal{L}(\lambda,x) \right] =
	\partial_x \mathcal{M}_n (\lambda,x) + :\left[ \mathcal{M}_n(\lambda,x),\mathcal{L}(\lambda,x)\right] \label{sklyaninMmatrixrelation}:
\end{equation}
for the first two charges: the number operator $I_0 = \mathcal{N}$ (\ref{psinumber}) and the momentum operator $I_1 = \mathcal{P}$ (\ref{psimomentum}) with $\mathcal{M}_0(\lambda,x) = \nicefrac{i \sigma_3}{2}$ and $\mathcal{M}_1(\lambda,x) = - \mathcal{L} (\lambda,x)$, where $:\ldots:$ denotes the normal ordering with respect to the fields $\Psi_n(x)$ \cite{Sklyanin:1988s1}. The relation (\ref{sklyaninMmatrixrelation}) yields:
\begin{equation}
	\label{sklyaninTMmatrixrelation} i \left[I_n, T^{x_+}_{x_-}(\lambda) \right] = :\mathcal{M} _n(\lambda,x_+)T^{x_+}_{x_-}(\lambda) - T^{x_+}_{x_-}(\lambda)\mathcal{M} _n(\lambda,x_-):
\end{equation}
from which the commutation relations follow\footnote{ Here we consider these relations in the isotropic limit.}:
\begin{eqnarray}
	\left[ \mathcal{N}, \log A(\lambda) \right] = 0 &,& \left[\mathcal{N}, B(\lambda) \right] = B(\lambda) \label{SklyaninNcommutators} \\
	\left[ \mathcal{P}, \log A(\lambda) \right] = 0 &,& \left[\mathcal{P}, B(\lambda) \right] = -\frac{2}{\lambda} B(\lambda) \label{SklyaninPcommutators}
\end{eqnarray}

In \cite{Sklyanin:1988s1} the action of the quantum-mechanical Hamiltonian on the two-particle state, as well as the commutation relations (in the infinite interval limit):
\begin{equation}
	\left[ \mathcal{H}, \log A(\lambda) \right] = 0 \quad , \quad \left[\mathcal{ \ H}, B(\lambda) \right] = \frac{4}{\lambda^2}B(\lambda) \label{SklyaninHcommutators}
\end{equation}
leading to the spectrum, were conjectured, so that they reproduce the correct classical limit. In the next section we will address this issue in details, and the relations (\ref{SklyaninHcommutators}) will be properly derived. Thus, the importance of the $M$-operators here is to establish a connection between the inverse scattering method, the direct diagonalization, and self-adjointness of the correct local conserved charges.

\section{Operator regularization} \label{operatorregularization}

In this section we propose a method to deal with all singularities associated with the operator product at the same point, and derive a number of results as a consequence. In particular, we will obtain the quantum analogue for the three fundamental invariants (\ref{invariants}), derive their spectrum and the corresponding $\mathcal{M}_{m}(\lambda;x,t)$ operators. We will also explain in which sense this method can be considered as an alternative to the discretization procedure.

As previously discussed, the most challenging problem of the LL-model, as well as all continuous integrable models, is the difficulty to express the commuting quantities as integrals of local densities. The root of this problem can be traced back to the fact that such local charges contain the product of operators at the same point, which in their turn make the aforementioned integrals very ill-defined quantities. Our solution to this problem is to deal from the very beginning with a set of regularized fields.
\begin{equation}
	\label{regularizedS} S^{i}_{\mathcal{F}} (x) := \int d\xi\: \mathcal{F} _{\epsilon}(x,\xi) S^{i}(\xi)\:, \quad i=+,-,3
\end{equation}
where $\mathcal{F}_{\epsilon}(x,y)$ is some smooth, rapidly decreasing, symmetric function of $(x,y)$, depending on some parameter $\epsilon$ that renders (\ref{regularizedS}) well-defined. More conditions on this $\mathcal{ \ F}_{\epsilon}(x,y)$ will be imposed later. Our goal will be a complete reformulation of the singular theory (which involves the bare $S$-fields) in terms of the regularized $S_{\mathcal{F}}^{i}(x)$-fields.

It is important to emphasize that the introduction of such a function is more in the spirit of the $F_{\epsilon}(x,y)$ introduced by \cite {Melikyan:2008ab}\footnote{Although the function $F_{\epsilon}(x,y)$ introduced in \cite{Melikyan:2008ab} differs from the function $\mathcal{F}_{\epsilon}(x,y)$ we use in the definition of (\ref{regularizedS}), the connection between the two is given, as will become obvious from the text below, by the expression: $F_{\epsilon}(x,y)\sim\int d\alpha \mathcal{F_{\epsilon}}(x,\alpha) \mathcal{F}_{\epsilon}(\alpha,y)$. Although it was not important for the purposes of \cite {Melikyan:2008ab}, it will become clear that it is necessary to introduce here the $\mathcal{F}_{\epsilon}(x,y)$ function to deal with singularities in the algebra and the Yang-Baxter relation. This will effectively affect the scaling of the operators, and result in correct $su(1,1)$ representation.} to regularize the quantum Hamiltonian (see equation (\ref{splitpointhamiltonian})), and that it differs in its essence from the approach employed by Sklyanin in \cite{Sklyanin:1988s1}, where the product of operators at the same point was modified by the introduction of the $\circ$-product, defined in (\ref{sklyaninproduct}). Here, the operator product is kept unchanged, i.e., it is still the usual operator product, while the operators themselves are regularized according to (\ref{regularizedS}), and, as it will become clear below, also renormalized. In what follows, we will show that the introduction of such regularization allows us to satisfy the bilinear relation (\ref{bilinearrelation}) without the need to resort to the product (\ref{sklyaninproduct}).

As a first step in this direction we obtain the algebra satisfied by the regularized $S^{i}_{\mathcal{F}}(x)$. Inverting the relation (\ref{regularizedS}),
\begin{equation}
	S^{i}(\xi) = \int dz\: \mathcal{G}_{\epsilon}(\xi,z) S^{i}_{ \mathcal{F}}(z)
\end{equation}
we introduce $\mathcal{G}_{\epsilon}(x,y)$ a smooth, rapidly decreasing, symmetric function of $(x,y)$, depending on $\epsilon$ and such that:
\begin{equation}
	S^{i}_{\mathcal{F}}(x) = \int d\xi \: dz\: \mathcal{F} _{\epsilon}(x,\xi) \mathcal{G}_{\epsilon}(\xi,z) S^{i}_{\mathcal{F}}(z) \quad \Leftrightarrow \quad \int d\xi \: \mathcal{F}_{\epsilon}(x,\xi) \mathcal{G}_{\epsilon}(\xi,z) = \delta(x-z)
\end{equation}

Now, with the use of the algebra satisfied by the bare operators (\ref{canonicalcommutationrelations}), one easily derives the algebra for the regularized $S_{\mathcal{F}}$-operators:
\begin{align}
	\label{regularizedalgebra} \left[ S^3_{\mathcal{F}}(x); S^{\pm}_{\mathcal{F}}(y) \right] &= \pm \int dz\: \mathcal{K}_{\epsilon}(x,y,z) S^{\pm}_{\mathcal{F}}(z) \notag \\
	\\
	\left[ S^-_{\mathcal{F}}(x); S^+_{\mathcal{F}}(y) \right] &= 2 \int dz\: \mathcal{K}_{\epsilon}(x,y,z) S^{3}_{\mathcal{F}}(z) \notag
\end{align}
where the kernel $\mathcal{K}_{\epsilon}(x,y,z)$ is the symmetric function of $(x,y)$ defined as follows:
\begin{equation}
	\label{kernelforregularizedSoperators} \mathcal{K}_{\epsilon}(x,y,z) \equiv \int d\xi\: \mathcal{F} _{\epsilon}(x,\xi)\mathcal{F}_{\epsilon}(y,\xi)\mathcal{G}_{\epsilon}(z,\xi)
\end{equation}

Let us note, that although the algebra of the regularized operators is non-local, it is however local after removing the regularization $\epsilon \longrightarrow 0$ (see the Eq. (\ref{F}) below). Having obtained the algebra for the $S_{\mathcal{F}}$-operators, we return to the bilinear relation (\ref{bilinearrelation}), which we now write in terms of the regularized operators $S_{\mathcal{F}}^i$. As discussed previously, the conditions needed to ensure the validity of (\ref{bilinearrelationforL}) are of the form:
\begin{equation}
	\label{conditionsforFYB} \left[ S^3_{\mathcal{F}}(x); S^{\pm}_{\mathcal{F}}(x) \right] = \pm S^{\pm}_{ \mathcal{F}}(x) \quad \mathrm{and} \quad \left[ S^-_{\mathcal{F} }(x); S^{+}_{ \mathcal{F}}(x) \right] = 2 S^{3}_{\mathcal{F}}(x)
\end{equation}
Thus, as long as the algebra (\ref{conditionsforFYB}) is satisfied, the relation (\ref{bilinearrelation}) follows. It is easy to show that this requirement leads to the following constraint on the $\mathcal{F}_{\epsilon}(x,y)$ function:
\begin{equation}
	\lim_{\epsilon \to 0}\left[ \mathcal{F}_{\epsilon}(x) \right]^2 = \lim_{\epsilon \to 0} \mathcal{F}_{\epsilon}(x) \label{FFF}
\end{equation}
We can conclude that with this restriction on the $\mathcal{F} _{\epsilon}(x,y)$ function in the $\epsilon \to 0$ limit, the algebra (\ref{regularizedalgebra}) reduces to the conditions (\ref{conditionsforFYB}) in the limit $x=y$, $\epsilon \to 0$ and, therefore, the relation (\ref{bilinearrelationforL}) is valid in the $\epsilon \to 0$ limit, with $\circ$ taken as the usual operator product, but with the $\mathcal{L}$ -operator defined in terms of the regularized $S_{\mathcal{F}}$-fields, i.e.,
\begin{align}
	\lim_{\epsilon \to 0} \left\{ R(\lambda_1-\lambda_2) \left[ \overset{(1)}{ \mathcal{L}}_{\mathcal{F}}(\lambda_1,x) + \overset{(2)}{\mathcal{L}}_{ \mathcal{F}}(\lambda_2,x) + \overset{(1)}{\mathcal{L}}_{\mathcal{F} }(\lambda_1,x) \cdot \overset{(2)}{\mathcal{L}}_{\mathcal{F}}(\lambda_2,x) \right] \right\} = \notag \\
	= \lim_{\epsilon \to 0} \left\{ \left[ \overset{(1)}{\mathcal{L}}_{\mathcal{ F }}(\lambda_1,x) + \overset{(2)}{\mathcal{L}}_{\mathcal{F}}(\lambda_2,x) + \overset{(2)}{\mathcal{L}}_{\mathcal{F}}(\lambda_2,x) \cdot \overset{(1)}{ \mathcal{L}}_{\mathcal{F}}(\lambda_1,x) \right] R(\lambda_1-\lambda_2) \right\} \label{bilinearrelationforLF}
\end{align}
and
\begin{equation}
	\label{LF-operator} \mathcal{L}_{\mathcal{F}}(\lambda,x) = \frac{i}{\lambda}\left(
	\begin{array}{cc}
		S^3_{\mathcal{F}}(x) & -S^+_{\mathcal{F}}(x) \\
		S^-_{\mathcal{F}}(x) & -S^3_{\mathcal{F}}(x)
	\end{array}
	\right)
\end{equation}
But now, we can use (\ref{bilinearrelationforLF}) to ensure the validity of the bilinear relation (\ref{bilinearrelation}) in the limit $\epsilon \to 0$, with the monodromy matrix (\ref{monmat}) defined with respect to the regularized $\mathcal{L}_{\mathcal{F}}$-operator. As explained before, the Eq. (\ref{bilinearrelation}) is fundamental for the applicability of the quantum inverse scattering method. Although the requirement (\ref{FFF}) does not have a non-trivial solution in the class of usual functions, it does have a solution in the class of generalized functions, and can be written as
\begin{equation}
	\mathcal{F_{\epsilon}}(x,y) = \frac{\delta_\epsilon (x-y)}{\delta_{\epsilon}(0)} \label{F}
\end{equation}
where $\delta_{\epsilon}(x)$ is some regularization of the $\delta$-function, so $\delta_{\epsilon}(0)$ is a well-defined constant before taking the limit $\epsilon \to 0$. Thus, the function $\mathcal{F}_{\epsilon}(x,y)$ is not only a regularization, but a renormalization as well. Hence, in the case of a product of two operators it will lead to a product like (\ref{sklyaninproduct}), but without its problematic, as previously discussed, behavior on regular fields.

\subsection {Casimir operator} \label{casop} One question that naturally arises in this context, is what happens in the quantum theory to the classical constraint (\ref{scalarproduct}):
\begin{equation}
	\label{casimir} \mathcal{C} \equiv S^3(x)S^3(x) - S^+(x)S^-(x) = 1
\end{equation}

It is not difficult to see that the quantum operator defined by $\mathcal{C}$ is formally a Casimir operator of the algebra (\ref{canonicalcommutationrelations}), but once again, due to the presence of the product of operators at the same point, it is clear that the expression (\ref{casimir}) is not well-defined. There are two possible ways to circumvent this problem. One can either consider the bare operators and replace the product by the Sklyanin $\circ$-product (\ref{sklyaninproduct}), or write (\ref{casimir}) using regularized $S_{\mathcal{F}}$ operators, but with usual product. Even though, we have already pointed out some of the advantages of using the regularized fields when compared to the product (\ref{sklyaninproduct}), we would like to work out this point in more detail for the product (\ref{sklyaninproduct}), so that one of the aforementioned problems of this product becomes clear.

We already saw (see the discussion after (\ref{prototypeofrelationssklyaniproduct})) that $\mathds{1} \circ \mathds{1} = 0$. Here we give an example of an even more serious inconsistency of $\circ$-product. Noting that:
\begin{align}
	S^3(x) \circ S^3(x) |0\rangle &= \lim_{\Delta \to 0} \frac{1}{\Delta} \int_x^{x+\Delta} d\xi_1 \: \int_x^{x+\Delta} d\xi_2 \: S^3(\xi_1) S^3(\xi_2) |0\rangle \notag \\
	& = \lim_{\Delta \to 0} \frac{1}{\Delta} \int_x^{x+\Delta} d\xi_1 \: \int_x^{x+\Delta} d\xi_2 \: |0\rangle = \lim_{\Delta \to 0} \frac{\Delta^2}{ \Delta} |0\rangle = 0
\end{align}
one arrives at meaningless result. Hence, we see that working directly with bare operators, even with the $\circ$-product (\ref{sklyaninproduct}), does not allow one to perform a completely consistent regularization of the model.

The second possibility is to use the regularized $S_{\mathcal{F}}^i$ operators. Let us begin with the general expression for a candidate for the Casimir operator:
\begin{equation}
	\label{regularizedconstraint} \mathcal{C}_{\mathcal{F}} = S^3_{\mathcal{F}}(x)S^3_{\mathcal{F}}(x) - S_{\mathcal{F}}^+(x)S_{\mathcal{F}}^-(x) + \gamma_{1} S_{\mathcal{F} }^3 = \kappa_{\epsilon}
\end{equation}
Here the parameters $\gamma_1$ and $\kappa_{\epsilon}$ are introduced to reflect the fact that in the quantum theory the operators $S^+(x)$ and $S^-(x)$ do not commute. We also note, that now both sides depend on the regularization parameter $\epsilon$, and, although the expression (\ref{regularizedconstraint}) is not a Casimir of the algebra (\ref{regularizedalgebra}) anymore, it can be checked that it becomes one in the limit $\epsilon \to 0$. Thus, more care is needed when using the quantum regularized constraint (\ref{regularizedconstraint}), and, in general, one cannot assume that $\mathcal{C} = 1$, as in the classical case, since we use rescaled operators. We will now determine the parameter $\gamma_1$, and in subsequent sections also the parameter $\kappa_{\epsilon}$. In the $\epsilon \to 0$ limit it will be shown that we indeed constructed spin $s=\nicefrac{1}{2}$ representation from the $S_{\mathcal{F}}$ fields.

Using the $\mathcal{L}_{\mathcal{F}}$-operator (\ref{LF-operator}) and the algebra (\ref{regularizedalgebra}), one finds (in the $\epsilon\rightarrow 0$ limit):
\begin{equation}
	(\mathcal{L}_{\mathcal{F}})^2 = \frac{i}{\lambda}- \mathds{1} \left( \kappa_\epsilon - (\gamma_1 +1)S_{\mathcal{F}}^3 \right) \label{Ltwo}
\end{equation}
Then, for $(\mathcal{L}_{\mathcal{F}})^3$ to be uniquely defined, i.e. so that $(\mathcal{L}_{\mathcal{F}})^3=(\mathcal{L}_{\mathcal{F}})\cdot(\mathcal{L}_{\mathcal{F}})^2=(\mathcal{L}_{\mathcal{F}})^2\cdot (\mathcal{L}_{\mathcal{F}})$, one must impose the condition $[\mathcal{L}_{\mathcal{F}},S_{\mathcal{F}}^3]=0$, which leads to the determination of the $\gamma_1$ parameter:
\begin{equation}
	\gamma_1 = -1
\end{equation}
Thus, the regularized constraint (\ref{regularizedconstraint}) takes the form:
\begin{align}
	\mathcal{C}_{\mathcal{F}} = S^3_{\mathcal{F}}(x)S^3_{\mathcal{F}}(x) -\frac{1}{2} \left[ S_{\mathcal{F}}^+(x)S_{\mathcal{F}}^-(x) +S_{\mathcal{F}}^-(x)S_{\mathcal{F}}^+(x) \right] = \kappa_{\epsilon} \label{regconst2}
\end{align}
We also note the following useful expression, which follows from (\ref{Ltwo}), and which will be used in the subsequent section when deriving the corresponding $M$ operator:
\begin{equation}
	[[\mathcal{L}_{\mathcal{F}},
	\partial_{x} \mathcal{L}_{\mathcal{F}}],\mathcal{L}_{\mathcal{F}}] =
	\partial_{x}\left[ \frac{1+4\kappa_{\epsilon}}{\lambda^2} \mathcal{L}_{\mathcal{F}} \right] \label{things_made_simpler}
\end{equation}

Finally, we renormalize the ferromagnetic vacuum in the following way. Let us define the ferromagnetic vacuum with the bare $S^3$ operator as before:
\begin{align}
	S^3|0\rangle = a_{0}|0\rangle \:, \quad S^- |0\rangle = 0
\end{align}
which in terms of the $S_{\mathcal{F}}^{3}$ operator takes the form:
\begin{equation}
	\label{regularizedvacuum} S^3_{\mathcal{F}} |0\rangle = \zeta_{\epsilon} |0\rangle \:, \quad S^-_{\mathcal{F}} |0\rangle = 0
\end{equation}
where the parameters $a_0$ and $\zeta_{\epsilon}$ will be determined at a later stage. Here we only notice, that using the definition (\ref{regularizedS}), and taking the limit $\epsilon \to 0$, the relation between the two parameters should be of the form:
\begin{equation}
	\zeta_{\epsilon} = \frac{a_0}{\delta_{\epsilon}(0)} \label{a_zeta}
\end{equation}
and from the Casimir operator (\ref{regularizedconstraint}) we obtain also the relation between the $\zeta_{\epsilon}$ and $\kappa_{\epsilon}$ parameters:
\begin{align}
	\zeta_{\epsilon}^{2} - \zeta_{\epsilon} = \kappa_{\epsilon} \label{zetakappa}
\end{align}

\subsection{$\mathcal{F}$-regularized Hamiltonian}

With the regularized $S_{\mathcal{F}}$-fields, all the products are now well defined, and we can proceed to the construction of the local conserved charges. The first step is to rewrite the quantum Hamiltonian (\ref{splitpointhamiltonian}) in terms of the $\mathcal{F}$-regularized fields:
\begin{equation}
	\label{fregularizedhamiltonian} \mathcal{H} = \frac{\eta(\epsilon)}{4} \int dx \: \left[ -2
	\partial_x S^3_{\mathcal{F} }(x)
	\partial_x S^3_{\mathcal{F}}(x) +
	\partial_x S^+_{\mathcal{F}}(x)
	\partial_x S^-_{\mathcal{F}}(x) +
	\partial_x S^-_{\mathcal{F}}(x)
	\partial_x S^+_{\mathcal{F}}(x) \right]
\end{equation}
where $\eta(\epsilon)$ is an arbitrary parameter which will be chosen later\footnote{In fact, one could start from a more general form: 
\begin{equation}
        \mathcal{H} = \frac{1}{4} \int dx \: \left[ -2 \eta_1(\epsilon)
	\partial_x S^3_{\mathcal{F} }(x)
	\partial_x S^3_{\mathcal{F}}(x) + \eta_2(\epsilon)
	\partial_x S^+_{\mathcal{F}}(x)
	\partial_x S^-_{\mathcal{F}}(x) + \eta_3(\epsilon)
	\partial_x S^-_{\mathcal{F}}(x)
	\partial_x S^+_{\mathcal{F}}(x) \right] \notag
\end{equation}
where the parameters $\eta_i(\epsilon)$, $i=1,2,3$ are arbitrary. However, the diagonalization condition will require setting $\eta_1(\epsilon)=\eta_2(\epsilon)=\eta_3(\epsilon) \equiv \eta(\epsilon)$, thus, we obtain the Hamiltonian in the form (\ref{fregularizedhamiltonian}).}. We now show that the same $n$-particle states of the form
\begin{equation}
	\label{fntilde} |f_n\rangle = \int \prod_{i=1}^n dx_i \: f _n(x_1,\ldots, x_n ) \prod_{j=1}^n S^+ |0\rangle
\end{equation}
provide a representation space for the $su(1,1)$ algebra for the operators in terms of $S_{\mathcal{F}}$ fields. As before, the $f_n(x_1,\ldots,x_n)$ are continuous and sufficiently fast decreasing, symmetric functions of $\{x_n\}$ for the integral (\ref{fntilde}) to be well-defined. The action of (\ref{fregularizedhamiltonian}) on the ferromagnetic vacuum gives the vacuum energy:
\begin{equation}
	\label{HFvacuum} \mathcal{H} |0\rangle = \frac{\eta(\epsilon)\zeta(\epsilon)}{2} \int dx\: dz\:
	\partial_x
	\partial_z \mathcal{K}_{\epsilon}(x,y,z) \vert_{x=y} |0\rangle \equiv h_0(\epsilon) |0\rangle
\end{equation}
while the action on the one-particle state gives:
\begin{equation}
	\label{HFoneparticle} (\mathcal{H} - h_0(\epsilon) )|f_1\rangle \overset{\epsilon \to 0}{ \longrightarrow} - \frac{a_{0}\eta(\epsilon)}{\delta^{2}(0)} \int dx\:
	\partial_x^2 f_1(x) S^+ (x)|0\rangle
\end{equation}
Here and below we use the symbol $\delta(0)$ for $\lim_{\epsilon\rightarrow 0} \delta_\epsilon (0)$. For the more complex two-particle sector, it is possible to show that:
\begin{align}
	(\mathcal{H} - h_{0}(\epsilon)) |f_2\rangle &\overset{\epsilon \to 0}{ \longrightarrow} - \frac{a_{0}\eta(\epsilon)}{\delta^{2}(0)} \int dx \: dy\: \triangle_2 f_2(x,y) S^+(x)S^+(y)|0\rangle + \notag \\
	&+ \eta(\epsilon)\left[ \frac{a_{0}}{\delta^{2}(0)}- \frac{1}{2\delta(0)}\right] \int dx\: \left[(
	\partial_x -
	\partial_y) f_2(x,y) \right]^{y=x+\epsilon}_{y=x- \epsilon}S^+(x)S^+(x)|0\rangle - \notag \\
	&- \frac{\eta(\epsilon)}{\delta^{2}(0)}\int dx\:
	\partial_x
	\partial_y f_2(x,y) \vert_{x=y}S^+(x)S^+(x)|0\rangle \label{HFtwoparticle}
\end{align}
Thus, choosing the old matching condition (\ref{matchingconditiossplitpointhamiltonian}) we find the value of the $a_0$ parameter:
\begin{align}
	a_0 = 1+\frac{\delta(0)}{2} \label{integrable}
\end{align}
It is easy to understand this result by considering the Casimir operator\footnote{We emphasize that it is a Casimir only in the $\epsilon \to 0$ limit.} (\ref{regularizedconstraint}). Remembering the relations (\ref{a_zeta}) and (\ref{zetakappa}) we obtain the $\kappa_{\epsilon}$ parameter in the $\epsilon \to 0$ limit:
\begin{equation}
	\mathcal{C}_{\mathcal{F}} = S^3_{\mathcal{F}}(x)S^3_{\mathcal{F}}(x) -\frac{1}{2} \left[ S_{\mathcal{F}}^+(x)S_{\mathcal{F}}^-(x) +S_{\mathcal{F}}^-(x)S_{\mathcal{F}}^+(x) \right] = -1/4 \label{kappa2}
\end{equation}
Thus, the constructed $su(1,1)$ representation indeed corresponds to spin $s=\nicefrac{1}{2}$. This is because the $S_{\mathcal{F}}^{i}$ operators were introduced in such a way that at each fixed point $x$, the $su(1,1)$ algebra they form in the $\epsilon \to 0$ limit (\ref{conditionsforFYB}) is well defined. This is unlike the discretization procedure, where one needs to introduce discrete $S_n$ operators and construct non-integrable (i.e., $s \neq \nicefrac{1}{2}, 1, etc.$) representations.

We can now also fix the value of $\eta(\epsilon)$, by choosing its limiting behavior: $\eta(\epsilon) \to E_{0}\frac{\delta^2(0)}{a_{0}}$ when $\epsilon \to 0$, where $E_{0}$ is an arbitrary {\em finite} constant. Thus, the expressions (\ref{HFoneparticle}) and (\ref{HFtwoparticle}) take the form:
\begin{align}
	&(\mathcal{H} - h_0(\epsilon) )|f_1\rangle \overset{\epsilon \to 0}{ \longrightarrow} - E_{0} \int dx\:
	\partial_x^2 f_1(x) S^+(x)|0\rangle \notag \\
	&(\mathcal{H} - h_{0}(\epsilon)) |f_2\rangle \overset{\epsilon \to 0}{ \longrightarrow} - E_{0}\int dx \: dy\: \triangle_2 f_2(x,y) S^+(x)S^+(y)|0\rangle \label{renormham2}
\end{align}

Consequently, it is possible to reproduce the construction of the self-adjoint extensions for the two-particle sector of \cite{Melikyan:2008ab}, as discussed in the end of the Section \textbf{\ref{overviewofLLquant}}. More generally, it is possible to show that that relations similar to (\ref{matchingconditiossplitpointhamiltonian}) and (\ref{splitpointhamiltonianenergy}) hold for the $\mathcal{F}$-regularized Hamiltonian (\ref{fregularizedhamiltonian}), so that the general construction of the self-adjoint extension of \cite{Melikyan:2008ab} also goes through in the $n$-particle sector for (\ref{fregularizedhamiltonian}).

Finally, we note that the regularized Hamiltonian (\ref{fregularizedhamiltonian}) can now be written in the simple form, using the trace and $\mathcal{L} _{\mathcal{F}}$ operators as follows:
\begin{equation}
	\mathcal{H} =h_0 + \frac{\lambda^2}{4} \int dx\: \eta(\epsilon){\tr} \left[
	\partial_x \mathcal{L} _{\mathcal{F}}(\lambda,x) \cdot
	\partial_x \mathcal{L}_{\mathcal{ F} }(\lambda,x) \right] \label{ham_LF}
\end{equation}
This would not have been possible without the introduction of the regularized fields. Thus, our first result is that the form of the quantum regularized Hamiltonian (\ref{ham_LF}) essentially coincides with the classical fundamental invariant in (\ref{invariants}), with classical fields replaced by the $S_{\mathcal{F}}^{i}$ operators. Below we will also obtain the other higher-order fundamental invariant in terms of the $S_{\mathcal{F}}^{i}$ fields, and show that it has the same form as its classical counterpart. Moreover, it will also be written in the form of the trace of $\mathcal{L} _{\mathcal{F}}$-operators.

\subsection{Spectrum}

In the previous section we have found the well-defined quantum Hamiltonian (\ref{ham_LF}) and derived its spectrum by means of the direct diagonalization. In this section we show that the spectrum can be also obtained from (\ref{ham_LF}) by means of the inverse scattering method. We establish a connection between the two methods by deriving the commutation relations (\ref{SklyaninHcommutators}).

The first step consists of finding the operator $\mathcal{M} _2(\lambda,x)$ (see Section \textbf{(\ref{overviewofLLMoperator})}) which satisfies the relation:
\begin{equation}
	\label{M2matrixrelation} i \left[ \mathcal{H}; \mathcal{L}_{\mathcal{F}}(\lambda,x) \right] =
	\partial_x \mathcal{M}_2(\lambda,x) + \left[\mathcal{M}_{2}(\lambda,x); \mathcal{L}_{\mathcal{F}}(\lambda,x) \right]
\end{equation}
It is not difficult to compute the left hand side of (\ref{M2matrixrelation}):
\begin{equation}
	\label{commutatorhflf} i [\mathcal{H},\mathcal{L}_{\mathcal{F}}(\lambda,x)] \overset{\epsilon \to 0} {\longrightarrow} -\frac{\eta(\epsilon)\lambda}{2\delta_{\epsilon}(0)}
	\partial_x [ \mathcal{L}_{\mathcal{F}}(\lambda,x),
	\partial_x \mathcal{L}_{\mathcal{F} }(\lambda,x)]
\end{equation}
Now, using the relation (\ref{things_made_simpler}) it is easy to check that the $\mathcal{M} _2(\lambda,x)$ operator has the form:
\begin{equation}
	\label{m2ansatz} \mathcal{M}_2(\lambda,x) = \alpha_2 \mathcal{L}_{\mathcal{F}}(\lambda,x) + \beta_2 [\mathcal{L}_{\mathcal{F}}(\lambda,x),
	\partial_x \mathcal{L}_{ \mathcal{F}}(\lambda,x)]
\end{equation}
where the $\alpha_2$ and $\beta_2$ coefficients have the form:
\begin{equation}
	\alpha_2 = \frac{1+4\kappa_{\epsilon}}{2\lambda \delta_{\epsilon}(0)}\:, \quad \beta_2 = - \frac{\lambda}{2 \delta_{\epsilon}(0)} \label{alphabet}
\end{equation}
Finally, from the equation (\ref{M2matrixrelation}) one obtains:
\begin{equation}
	\label{M2matrixtransferfinite} i \left[ \mathcal{H}; T^{x_+}_{x_-}(\lambda)\right] = \mathcal{M} _2(\lambda,x_+) T^{x_+}_{x_-}(\lambda) - T^{x_+}_{x_-}(\lambda) \mathcal{M} _2(\lambda,x_-)
\end{equation}
where the transfer matrix is defined with respect to $\mathcal{L}_{\mathcal{F}}$, and decomposing the $\mathcal{M}_2(\lambda,x)$ as:
\begin{equation}
	\label{M2decomposition} \mathcal{M}_2(\lambda,x) = \sigma_+ \mathcal{M}_2^+(\lambda,x) + \sigma_- \mathcal{M}_2^-(\lambda,x) +\sigma_3 \mathcal{M}_2^3(\lambda,x)
\end{equation}
where:
\begin{align}
	\mathcal{M}_2^+(\lambda,x) &= - \frac{i \alpha_2}{\lambda} S^+_{\mathcal{F} }(x) - \frac{2 \beta_2}{\lambda^2} \left[ S^+_{\mathcal{F}}(x)
	\partial_x S^3_{\mathcal{F}}(x) -
	\partial_x S^+_{\mathcal{F}}(x) S^3_{\mathcal{F}}(x) \right] \notag \\
	\mathcal{M}_2^-(\lambda,x) &= \frac{i \alpha_2}{\lambda} S^-_{\mathcal{F} }(x) + \frac{2 \beta_2}{\lambda^2} \left[ S^3_{\mathcal{F}}(x)
	\partial_x S^-_{\mathcal{F}}(x) -
	\partial_x S^3_{\mathcal{F}}(x) S^-_{\mathcal{F}}(x) \right] \label{M2decompositioncomponents} \\
	\mathcal{M}_2^3(\lambda,x) &= \frac{i \alpha_2}{\lambda} S^3_{\mathcal{F} }(x) + \frac{\beta_2}{\lambda^2} \left[ S^+_{\mathcal{F}}(x)
	\partial_x S^-_{ \mathcal{F}}(x) -
	\partial_x S^+_{\mathcal{F}}(x) S^-_{\mathcal{F}}(x) \right] \notag
\end{align}
the equation (\ref{M2matrixrelation}) is reduced to the set of commutation relations for the elements of the transfer matrix (\ref{monmat}):
\begin{align}
	\label{commutationrelationsforHandTfinite} i \left[ \mathcal{H}; A^{x_+}_{x_-}(\lambda)\right] &= \mathcal{M} _2^+(\lambda,x_+)C^{x_+}_{x_-}(\lambda) + \mathcal{M}_2^3( \lambda,x_+)A^{x_+}_{x_-}(\lambda) - A^{x_+}_{x_-}(\lambda) \mathcal{M} _2^3(\lambda,x_-) - B^{x_+}_{x_-}(\lambda) \mathcal{M}_2^-(\lambda,x_-) \notag \\
	i \left[ \mathcal{H}; B^{x_+}_{x_-}(\lambda)\right] &= \mathcal{M} _2^+(\lambda,x_+)D^{x_+}_{x_-}(\lambda) + \mathcal{M}_2^3( \lambda,x_+)B^{x_+}_{x_-}(\lambda) - A^{x_+}_{x_-}(\lambda) \mathcal{M} _2^+(\lambda,x_-) + B^{x_+}_{x_-}(\lambda) \mathcal{M}_2^3(\lambda,x_-) \notag \\
	\\
	i \left[ \mathcal{H}; C^{x_+}_{x_-}(\lambda)\right] &= \mathcal{M} _2^-(\lambda,x_+)A^{x_+}_{x_-}(\lambda) - \mathcal{M}_2^3( \lambda,x_+)C^{x_+}_{x_-}(\lambda) - D^{x_+}_{x_-}(\lambda) \mathcal{M} _2^-(\lambda,x_-) - C^{x_+}_{x_-}(\lambda) \mathcal{M}_2^3(\lambda,x_-) \notag \\
	i \left[ \mathcal{H}; D^{x_+}_{x_-}(\lambda)\right] &= \mathcal{M} _2^-(\lambda,x_+)B^{x_+}_{x_-}(\lambda) - \mathcal{M}_2^3( \lambda,x_+)D^{x_+}_{x_-}(\lambda) + D^{x_+}_{x_-}(\lambda) \mathcal{M} _2^3(\lambda,x_-) - C^{x_+}_{x_-}(\lambda) \mathcal{M}_2^+(\lambda,x_-) \notag
\end{align}

Passing to the infinite interval limit is accomplished by introducing the transfer matrix $T_{\infty}(\lambda)$ as the limit:
\begin{equation}
	T_{\infty}(\lambda) = \lim_{x_{\pm} \to \pm \infty} \left[ e(-x_+,\lambda) T^{x_+}_{x_-}(\lambda) e(x_-,\lambda) \right]
\end{equation}
where $e(x,\lambda) \equiv \exp\left(\nicefrac{i \sigma_3}{\lambda}x\right)$ . Then, by multiplying both sides of the equations (\ref{commutationrelationsforHandTfinite}) by $ e(-x_+,\lambda)$ from the left and $e(x_-,\lambda)$ from the right, taking the limit $x_{\pm} \to \pm \infty$ and noting that $S^3_{\mathcal{F}}(x) \overset{x \to \pm \infty}{\longrightarrow} 1$ and $S^{\pm}_{\mathcal{F}}(x) \overset{x \to \pm \infty}{\longrightarrow} 0$, one concludes that (in the limit $\epsilon \to 0$):
\begin{align}
	\left[ \mathcal{H}; A_{\infty}(\lambda)\right] &= 0 \\
	\left[ \mathcal{H}; B_{\infty}(\lambda)\right] &= \frac{2\alpha_2 \eta(\epsilon)}{ \lambda}B_{\infty}(\lambda) \\
	\left[ \mathcal{H}; C_{\infty}(\lambda)\right] &= -\frac{2\alpha_2 \eta(\epsilon)}{ \lambda}C_{\infty}(\lambda) \\
	\left[ \mathcal{H}; D_{\infty}(\lambda)\right] &= 0
\end{align}
Thus, we have proved the relations (\ref{SklyaninHcommutators}), from which the spectrum of the quantum Hamiltonian follows. This shows how to make a connection between the regularization scheme we have proposed, the direct diagonalization, and the inverse scattering method.

\subsection{Higher-order charges} \label{highercharges}

In this section we construct the quantum version of the cubic in fields invariant in (\ref{invariants}) and show that, similarly to the Hamiltonian, the density can be expressed in terms of the trace of a product of $\mathcal{L}_{\mathcal{F}}$-fields. We then conclude that since all three invariants (\ref{invariants}) have such form\footnote{It is obvious that the Casimir operator can also be written in such form $\mathcal{C}= \int dx\: {\tr} (\mathcal{L}_{\mathcal{F}} \mathcal{L}_{\mathcal{F}})$} and it coincides with their corresponding classical counterpart, any conserved polynomial should be expressed in terms of the three fundamental invariants, and, therefore can be written as a product of $\mathcal{L}_{\mathcal{F}}$-fields and their derivatives. Finally, based on this, we give a general prescription to obtain the corresponding $\mathcal{M}_m$ operator, which will establish the connection with the inverse scattering method.

Let us first give the final result for the cubic quantum regularized invariant $\mathcal{Q}_3$:
\begin{equation}
	\mathcal{Q}_3 = i\frac{c_0(\epsilon) \lambda^3}{2}\int dx\: {\tr} \left( \mathcal{L}_{\mathcal{F}}
	\partial_{x}{\mathcal{L}_{\mathcal{F}}}
	\partial_{x}^{2} {\mathcal{L}_{\mathcal{F}}} \right) \label{qq3}
\end{equation}
where the constant $c_0(\epsilon)$ will be fixed later. To derive it, one starts with a general cubic polynomial of the form:
\begin{align}
	\mathcal{Q}_3 = i c_0(\epsilon) \int dx\: & \left[ a_1S^{3}_{\mathcal{F}}
	\partial_{x} S^{+}_{\mathcal{F}}
	\partial_{x}^{2}S^{-}_{\mathcal{F}} + a_2 S^{3}_{\mathcal{F}}
	\partial_{x} S^{-}_{\mathcal{F}}
	\partial_{x}^{2}S^{+}_{\mathcal{F}} + a_3 S^{+}_{\mathcal{F}}
	\partial_{x} S^{3}_{\mathcal{F}}
	\partial_{x}^{2}S^{-}_{\mathcal{F}} \right. \label{q3_gen}\\ \notag
	& \\ \notag
	&\left. +a_4 S^{+}_{\mathcal{F}}
	\partial_{x} S^{-}_{\mathcal{F}}
	\partial_{x}^{2}S^{3}_{\mathcal{F}} +a_5 S^{-}_{\mathcal{F}}
	\partial_{x} S^{3}_{\mathcal{F}}
	\partial_{x}^{2}S^{+}_{\mathcal{F}} +a_6 S^{-}_{\mathcal{F}}
	\partial_{x} S^{+}_{\mathcal{F}}
	\partial_{x}^{2}S^{3}_{\mathcal{F}}\right] \notag
\end{align}
where the coefficients $a_{1}, ...,a_{6}$ are arbitrary and will be fixed from the diagonalization condition. It is easy to show, using the algebra (\ref{regularizedalgebra}), that an arbitrary permutation of the derivatives will lead to the same form (\ref{q3_gen}). Now, after some tedious but straightforward calculations, one can show that the vector (\ref{representationofthealgebra}) diagonalizes the $\mathcal{Q}_3$ operator (\ref{q3_gen}), provided two conditions: i) one has to fix the relative coefficients $a_1 =-1, a_2 =1, a_3 =1, a_4 =-1, a_5 =-1, a_6=1$; and ii) to cancel the boundary terms, one has to use the same matching condition as before (see (\ref{matchingconditiossplitpointhamiltonian})), as well as use the $a_0$ parameter (\ref{integrable}), which lead to an integrable representation. Finally, with the relative coefficients fixed, one can collect the terms into a compact form:
\begin{equation}
	\mathcal{Q}_3 = i c_0(\epsilon) \int dx\: \epsilon^{ijk} (S_{\mathcal{F}})_{i}
	\partial_{x} (S_{\mathcal{F}})_{j}
	\partial_{x}^{2} (S_{\mathcal{F}})_{k} \label{q3_comp}
\end{equation}
From here, one can easily find the result (\ref{qq3}), where we expressed $(S_{\mathcal{F}})_{i}$ fields in terms of the $\mathcal{L}_{\mathcal{F}}(\lambda,x)$-operators. Choosing the function $c_0(\epsilon)$ such that $c_0(\epsilon) \to Q_0 \frac{\delta^2(0)}{4 a_{0}\zeta(\epsilon)}$ when $\epsilon \to 0$, where $Q_0$ is an arbitrary finite constant, one can show that, for example, for the two-particle sector:
\begin{align}
	\mathcal{Q}_3 |f_2\rangle \overset{\epsilon \to 0}{ \longrightarrow} - i Q_{0} \int dy_1 dy_2 \: \left(
	\partial_{y_{1}}^{3}+
	\partial_{y_{2}}^{3}\right) f_2(y_1,y_2) S^+(y_1)S^+(y_2)|0\rangle \label{q3spect}
\end{align}

Collecting the results of the previous sections, namely i) all three quantum fundamental invariants have the same form as their classical counterparts, when expressed in terms of the regularized operators; and ii) the three invariants can be expressed in terms of traces of $\mathcal{L}_{\mathcal{F}}(\lambda,x)$-operator product, we can now claim that an arbitrary conserved charge can be decomposed in terms of these invariants and their derivatives.

We now give a general prescription of obtaining the corresponding $\mathcal{M}_m$-operators, and, thus, making a connection to inverse scattering method. Since an arbitrary charge, as we have shown, can be expressed in terms of the $\mathcal{L}_{\mathcal{F}}(\lambda,x)$-operator product, it is convenient to use the relation:
\begin{equation}
	\left[ \overset{(1)}{\mathcal{L}_{\mathcal{F}}}(x,\lambda_{1}), \overset{(2)}{\mathcal{L}_{\mathcal{F}}}(y,\lambda_{2})\right] = \int dz\: K_{\epsilon}(x,y,z)\left[ r(\lambda_{1},\lambda_{2}), \overset{(1)}{\mathcal{L}_{\mathcal{F}}}(z,\lambda_{1}) + \overset{(2)}{\mathcal{L}_{\mathcal{F}}}(z,\lambda_{2})\right] \label{r_mat}
\end{equation}
where the $r$-matrix has the form:
\begin{equation}
	r(\lambda_{1}, \lambda_{2}) = \frac{i}{2(\lambda_1 - \lambda_2)} \left(\mathds{1} \otimes \mathds{1} + \sum_{a} \sigma_{a} \otimes \sigma_{a} \right) \label{rmat2}
\end{equation}
Note, that in all final expressions, in the $\lambda_{1} \to \lambda_{2} = \lambda$ limit, the singularity $\nicefrac{1}{(\lambda_{1}-\lambda_{2})}$ will cancel. Then, for example, for the Hamiltonian, one finds:
\begin{equation}
	\left[ \mathcal{H}, {\mathcal{L}_{\mathcal{F}}}(y,\lambda_2)\right] = {\tr}_{(1)} \int dx\: \left[ \overset{(1)}{\mathcal{H}}(x,\lambda_{1}), \overset{(2)}{\mathcal{L}_{\mathcal{F}}}(y,\lambda_{2})\right] \label{pol1}
\end{equation}
which in the limit $\epsilon \to 0$ and $\lambda_{1} \to \lambda_{2}$ gives:
\begin{equation}
	i \left[ \mathcal{H}, \mathcal{L}_{\mathcal{F}}(y,\lambda)\right] = -\frac{\lambda \eta(\epsilon)}{2 \delta_{\epsilon}(0)}
	\partial_{y} \left[ \mathcal{L}_{\mathcal{F}},
	\partial_{y} \mathcal{L}_{\mathcal{F}} \right]
\end{equation}
leading to the previous result (\ref{M2matrixrelation}-\ref{alphabet}).

Similarly, using (\ref{r_mat}), one can derive the corresponding $\mathcal{M}_m$ operator for any charge, even though it becomes very tedious with each higher order. For example, for $\mathcal{Q}_3$, a similar calculation results in the following expression:
\begin{equation}
	\left[ \mathcal{Q}_{3},\mathcal{L}_{\mathcal{F}}(y,\lambda)\right] =\frac{i}{2}
	\partial_{y} \left[ \Lambda(\mathcal{L}_{\mathcal{F}}) \right]
\end{equation}
where $\Lambda(\mathcal{L}_{\mathcal{F}})$ is some function of the operator $\mathcal{L}_{\mathcal{F}}(y,\lambda)$.

\section{Self-adjoint operators and extensions}
\label{sec:self}

The final check of our construction is the verification of the self-adjointness of each operator. The self-adjointess of the Hamiltonian, together with the construction of self-adjoint extensions, was first worked out in \cite{Melikyan:2008ab} (see also Section {\textbf (\ref{overviewofLLquant}})). Here we will also generalize it to the next order fundamental invariant $\mathcal{Q}_3$.

Consider the scalar product (see \cite{Melikyan:2008ab}):
\begin{equation}
	\label{generalscalarproduct} \langle \tilde{g}_n | f_n \rangle = \sum_{\mathrm{Partitions}} \varepsilon^{n - M_P} C_P \int d^{M_P}t \tilde{g}_n^*(\mathbf{x}) f _n(\mathbf{x}) \vert_{\{x_i \in X_m\}=t_m}
\end{equation}
where $\{X_m\}$ is a partition of the set $\{x_i\}$:
\begin{equation}
	\bigcup_{m=1}^{M_P} X_m = \{x_i\} \quad \mathrm{and} \quad X_m \bigcap X_n = \delta_{mn} X_m \notag
\end{equation}
$t_m$ is a `collective' coordinate for all $x_i \in X_m$ and $C_P$ are positive combinatorial factors. We note that the case with $n=2$ is equivalent to the scalar product (\ref{scalarproductinV}). We consider here in details the simpler two-particle sector with the scalar product (\ref{scalarproductinV}), and in the end comment on the general $n$-particle case. Introducing the action of $ \mathcal{Q}_3$ on $|f_2\rangle$ as:
\begin{equation}
	\mathcal{Q}_3 | f_n \rangle = \left(
	\begin{array}{c}
		\hat{i}_3 f_1(x) \\
		- \triangle_3 f_2(x,y)
	\end{array}
	\right)
\end{equation}
with $\triangle_3 = \omega \sum_{i=1}^n
\partial_i^3$, where $n$ denotes the appropriate dimension, which in this case is $n=2$ and $\omega \in i \mathbb{R}$. Self-adjointness with respect to (\ref{scalarproductinV}), $\langle \mathcal{Q}_3 \tilde{g}_2 | f_n \rangle = \langle \tilde{g}_2 | \mathcal{Q}_3 f_n \rangle$, demands
\begin{align}
	\label{selfadjointnessofI3} \frac{1}{2} \int dx \: \tilde{g}^*_1(x) \hat{i}_3 f_1(x) &- \int_{x\neq y} dx\:dy\: \tilde{g}^*_2(x,y) \triangle_3 f_2(x,y) = \notag \\
	&= \frac{1}{2} \int dx \: \left(\hat{i}_3 \tilde{g}_1(x)\right)^* f _1(x) - \int_{x\neq y} dx\:dy\: \left(\triangle_3 \tilde{g}_2(x,y)\right)^* f_2(x,y)
\end{align}
Then, we compute
\begin{align}
	\int_{x\neq y} dx\: dy \: \tilde{g}_2^*(x,y) \triangle_3 f_2(x,y) &= \int_{x \neq y} dx\: dy\: \left( \triangle_3 \tilde{g}_2(x,y) \right)^* f_2(x,y) + \notag \\
	&+ \omega \int dx \: \left[ \tilde{g}^*_2(x,y) \left(
	\partial_x^2 -
	\partial_y^2 \right) f_2(x,y) + \left(
	\partial_x^2 -
	\partial_y^2 \right) \tilde{g}^*_2(x,y) f_2(x,y) \right. - \notag \\
	&- \left.
	\partial_x \tilde{g}_2^*(x,y)
	\partial_x f_2(x,y) +
	\partial_y\tilde{g}_2^*(x,y)
	\partial_y f_2(x,y) \right]_{y = x -\epsilon}^{y=x+\epsilon}
\end{align}
which is equivalent to
\begin{align}
	&\frac{3}{2} \int dx\: \left\{ \tilde{g}_2^*(x,y) \left[ \omega \left(
	\partial_x^2 -
	\partial_y^2 \right) \right] f_2(x,y) \right\}_{y=x-\epsilon}^{y=x+\epsilon} - \int_{x \neq y} dx\: dy\: \tilde{g} _2^*(x,y) \triangle_3 f_2(x,y) = \notag \\
	&\frac{3}{2} \int dx\: \left\{ \left[\omega \left(
	\partial_x^2 -
	\partial_y^2 \right) \tilde{g}_2(x,y)\right]^* f_2(x,y) \right\}_{y=x-\epsilon}^{y=x+\epsilon} - \int_{x \neq y} dx\: dy\: \left( \triangle_3 g_2(x,y) \right)^* f_2(x,y)
\end{align}
Comparison with (\ref{selfadjointnessofI3}) yields:
\begin{equation}
	\hat{i}_3 f_1(x) = 3 \left[ \omega \left(
	\partial_x^2 -
	\partial_y^2\right) f_2(x,y)\right]_{y=x-\epsilon}^{y=x+\epsilon} + \hat{i}_3^{\prime }f_1(x)
\end{equation}
where $\hat{i}_3^{\prime }$ is any self-adjoint operator in $\mathcal{L} ^2\left( \mathbb{R},dx\right)$. Next, the condition that the space generated by $| f_2\rangle$ is closed under the action of $\mathcal{Q}_3$ allows us to fix the form of $\hat{i}_3^{\prime }$.
\begin{align}
	-\triangle_3 f_2(x,y)\vert_{x=y} &= \hat{i}_3 f_1 \notag \\
	&= 3 \left[ \omega \left(
	\partial_x^2 -
	\partial_y^2\right) f _2(x,y) \right]_{y=x-\epsilon}^{y=x+\epsilon} + \hat{i}_3^{\prime }f _1(x) \notag \\
	\Rightarrow -\omega
	\partial_x^3 f_2(x,x) &= 3 \omega \left(
	\partial_x +
	\partial_y\right) \left[ \left(
	\partial_x -
	\partial_y\right) f_2(x,y) \vert_{y=x-\epsilon}^{y=x+\epsilon} -
	\partial_x
	\partial_y f_2(x,y)\vert|_{x=y} \right]+ \hat{i}_3^{\prime }f _1(x)
\end{align}
which upon the use of the matching condition (\ref{matchingconditiossplitpointhamiltonian}) for $n=2$:
\begin{equation}
	\left[
	\partial_x f_2(x,y) -
	\partial_y f_2(x,y) \right]_{y = x - \epsilon}^{y=x+\epsilon} =
	\partial_x
	\partial_y f _2(x,y)\vert_{x=y} \notag
\end{equation}
leads to the conclusion:
\begin{equation}
	\hat{i}_3^{\prime }f_1(x) = - \omega
	\partial_x^3 f_2(x,x)
\end{equation}
and finally, we are allowed to conclude that
\begin{equation}
	\mathcal{Q}_3 | f_n \rangle = \left(
	\begin{array}{c}
		3 \omega \left[\left(
		\partial_x^2 -
		\partial_y^2\right) f_2(x,y) \right]_{y=x-\epsilon}^{y=x+\epsilon} - \omega
		\partial_x^3 f_2(x,x) \\
		-\triangle_3 f_2(x,y)
	\end{array}
	\right) \label{seldadq3}
\end{equation}
is a self-adjoint operator. It is important to stress that (\ref{seldadq3}) does not impose any additional condition on the Hilbert space but rather is a consequence of the only matching conditions (\ref{matchingconditiossplitpointhamiltonian}) defining our Hilbert space. This result can be generalized for an arbitrary $n$-particle sector using the $S$-matrix factorization property (see \cite{Melikyan:2008ab} for details).

\section{Conclusion} \label{conclusion} In this paper we have proposed a method to quantize continuous integrable models on the concrete example of Landau-Lifshitz model, without resorting to discretization schemes. We also explained how to obtain the quantum trace identities, and established a connection with the inverse scattering method. The reason behind quantizing the system directly in the continuous framework is that discretization is not always an obvious procedure, and for more complicated models it is very desirable to learn to quantize integrable systems directly. The difficulty in quantizing the continuos model directly is the not well defined operator products, thus, leading to operator ordering problem. Even for the simplest models, such as the non-linear Schrödinger model, the normal ordering is a non-trivial procedure in quantum trace identities, and generally relies on some discretization procedure.

For the LL model the situation is more involved, since it is nor obvious how to sort three fields $S^3$, $S^+$ and $S^-$. One can try to resolve the constraint and use the normal ordering for the unconstrained fields, but this approach suffers from various problems, i.e., inability to find the quantum charges, etc. Besides, it is very desirable to obtain the quantum charges in terms of the original fields, namely, to obtain the quantum analogues of the classical expressions. In the theory with more than three fields, even after resolving the constraint, if it is even feasible, one will still face the problem of operator ordering. In this work we made the first steps to address all these problems. Although we work out these ideas on the particular LL model, and there are still several questions that need to be understood and generalized, we believe our method can be useful for quantizing other continuous integrable models.

Our method relies on several seemingly different aspects of the integrability, which we have shown to be inter-connected in the end. Firstly, the direct diagonalization requires two different procedures: i) operator regularization, and ii) construction of the extended Hilbert space. The operator regularization we have introduced is very general, and only satisfies a self-consistency condition following from the intertwining relation. This regularization, which at the same time was shown to renormalize the operators, has significant consequences in various aspects. Firstly, the algebra of the quantum operators was shown to be non-local even for the isotropic model. Let us remind, that previously such non-local algebra ({\em Sklyanin algebra}) appeared only in the anisotropic model. We also emphasize that our regularization is essentially different from the one originally introduced in \cite{Sklyanin:1988s1, Sklyanin:1982tf,Sklyanin:1979ll,Sklyanin:1983ig}. While in the original version the regularization (Sklyanin product) was chosen to satisfy the intertwining relation, it regularizes only one type of singularity. In the version we propose, the $\mathcal{F}$-regularization regularizes all singularities at the same time, and leads to well-defined operator products, as well as a well-defined algebra. We were able to show that this leads to integrable representations, unlike the discretization procedure, where the process of passing from discrete to continuous operators requires construction of non-integrable representations.

Secondly, the singular nature of the LL model (which is why one has to introduce operator regularization), requires the construction of self-adjoint extensions. While normally this is being ignored in physical models, it is nevertheless crucial for this model, as well as a number of other interesting models, and is relevant for strings in particular. To name a few simplest ones, the principal chiral model in the $R \times S^3$ subsector, the higher-order corrections (see \cite{Kruczenski:2004kw}), or the fermionic $AAF$ model will as well require the construction of self-adjoint extensions. One of the problems we would like to undertake is the nature of the minimal extension to accommodate the entire superstring on $AdS_5 \times S^5$. This is not an obvious question and will require a more detailed study of applications of functional analysis on superspace.

Together, the operator regularization, subject to quantum integrability, and the construction of self-adjoint extensions lead to several consequences: the S-matrix factorization, the nature of the representation, the connection to the inverse scattering method, etc. Let us once more stress on the latter point: the relation to inverse scattering method is established via the hierarchy of $\mathcal{M}_m$ operators. Namely, only after introducing the regularized operators and verifying the self-adjointness of the operators, it is possible to derive strictly the corresponding $\mathcal{M}_m$ operator. We have shown this directly on the examples of the quantum Hamiltonian and the higher-order cubic invariant. It was not possible to derive the $\mathcal{M}_2$ operator corresponding to the Hamiltonian before, while in our method, deriving the corresponding $\mathcal{M}_m$ operator for any conserved charge, although very tedious, is nevertheless very straightforward.

Finally, we have obtained the three fundamental invariants in the quantum theory such that, as in the classical theory, any conserved charge can be decomposed into these three invariants. Moreover, we have shown that all of them can be written in terms of the traces of the $\mathcal{L}_{\mathcal{F}}$ operator products and their space derivatives. This is exactly how one can obtain the $\mathcal{M}_m$ operators.

There are several problems one can undertake in the future. Even for the LL model, we have considered only the isotropic case. We have emphasized throughout the paper that anisotropy introduces not only technical difficulties, but is essentially different for the integrability. Namely, to satisfy the intertwining relations, one needs to introduce an additional quantum operator, thus, modifying the algebra of observables. This gives rise to quadratic Sklyanin algebras, which in turn involves the Sklyanin product extensively discussed in the text. It will be very interesting to investigate the anisotropic case using our prescription and shed some light on the origin of such quadratic algebras and their non-local nature. We expect that the anisotropy will completely change the extended Hilbert space and require more intense study of self-adjointness of the operators. We only mention the importance of the Sklyanin algebra,  as the discretization scheme, which leads to the {\emph local} charges in the discrete version, depends on representations of such algebras (see for example \cite{kundu1992ciq,tarasov1983lhi,tarasov1984lhi}). We hope to report on progress in this direction soon.

\section*{Acknowledgment} A.P. acknowledges partial support of CNPq under grant no.308911/2009-1. The work of G.W. was supported by the FAPESP grant No. 06/02939-9.

\bibliographystyle{utphys}
\bibliography{ll_charges_final_v2}

\end{document}